\documentclass[twocolumn,floatfix,superscriptaddress,preprintnumbers,nofootinbib,amsmath,amssymb,aps,prb]{revtex4-1}

\usepackage{graphicx}
\usepackage{booktabs}
\usepackage{float}        % improved float placement
\usepackage{placeins}     % \FloatBarrier via [section] (you had [section] option earlier)
\usepackage{multirow}
\usepackage{ragged2e}
\usepackage{tabularx}
\usepackage[table,xcdraw]{xcolor} % table colors
\usepackage{subcaption}
\usepackage{caption}

\usepackage{xfrac}
\usepackage[version=4]{mhchem}
\usepackage{chemformula}
\usepackage{siunitx}

\usepackage{lmodern}
\usepackage[T1]{fontenc}
\usepackage[utf8]{inputenc}
\usepackage[english]{babel}

\usepackage{epsfig}

\usepackage[
    colorlinks=true,
    linkcolor=blue,
    citecolor=blue,
    filecolor=blue,
    urlcolor=blue,
    allcolors=blue
]{hyperref}

\usepackage{todonotes}

\usepackage[
    font=small,
    labelfont=bf,
    justification=justified,
    format=plain % avoids hanging indentation
]{caption}
\usepackage{subcaption}

\usepackage{verbatim}
\usepackage{cleveref}
\usepackage{mathtools}
\usepackage{lipsum}   % useful for dummy text while writing

\usepackage[bottom]{footmisc}

\usepackage{wasysym}  % symbol set (you had it)
\usepackage{booktabs} % already above but safe to keep

\usepackage{tocloft}

\DeclareUnicodeCharacter{2212}{-}

\usepackage{booktabs}    % nicer horizontal rules
\usepackage{array}       % \extrarowheight
\usepackage{threeparttable}
\usepackage{siunitx}     % if you're using \sisetup / numeric alignment (optional)
\usepackage[version=4]{mhchem} % for \ce{}

\begin{document}

% ---------------- Authors and affilations ----------------
% authors.tex should contain the \author and \affiliation lines. Keep it separate
% collaborators can edit authorship without touching the structure of the paper.

% !TEX root = ./main.tex

\newcommand{\addressIFIMUP}{
	IFIMUP, Institute of Physics for Advanced Materials, Nanotechnology and Photonics, Departamento de Física e Astronomia da Faculdade de Ciências da Universidade do Porto, Rua do Campo Alegre, 687, 4169-007 Porto, Portugal
}

\newcommand{\addressIFUSP}{
	Universidade de São Paulo, Instituto de Física, Rua do Matão 1371, 05508-090, São Paulo, SP, Brazil
}

% -------------------------------------

\title{First-principles electric field gradients at A- and B-site cations across the NaRTiO$_{4}$ Ruddlesden–Popper series}

%% Author
\author{L. F. de Almeida}
\affiliation{\addressIFIMUP}

%% Author
\author{A. N. Cesário}
\affiliation{\addressIFIMUP}
%%
% Author
\author{P. A. Sousa}
\affiliation{\addressIFIMUP}
%
%% %%%%%%%%%%%%%%%%%%%%%%%%%%%%%%%%%%%%
%% Author
\author{P. Rocha-Rodrigues}
\affiliation{\addressIFIMUP}
%%
%% %%%%%%%%%%%%%%%%%%%%%%%%%%%%%%%%%%%%
%% Author
\author{L. V. C.  Assali}
\affiliation{\addressIFUSP}
%%
%% %%%%%%%%%%%%%%%%%%%%%%%%%%%%%%%%%%%%
%% Author
\author{H. M. Petrilli}
\affiliation{\addressIFUSP}
%%
%% %%%%%%%%%%%%%%%%%%%%%%%%%%%%%%%%%%%%
%% Author
\author{J. P. Araújo}
\affiliation{\addressIFIMUP}
%%
%% %%%%%%%%%%%%%%%%%%%%%%%%%%%%%%%%%%%%
%% Author
\author{A. M. L. Lopes}
\email{armandina.lopes@fc.up.pt}
\affiliation{\addressIFIMUP}
%%
%% %%%%%%%%%%%%%%%%%%%%%%%%%%%%%%%%%%%%

% ---------------- Date (optional) ----------------
% Uncomment the next line to always insert today's date.
% \date{\today}

% ---------------- Abstract ----------------
% abstract.tex includes the abstract.
 % !TEX root = ./main.tex
 
\begin{abstract}

\noindent The $n = 1$ Ruddlesden-Popper titanates, NaRTiO$_{4}$ (R = rare-earth), exhibit a structural behaviour where non-centrosymmetry is driven by cooperative oxygen octahedral rotations (OORs) rather than conventional second-order Jahn-Teller distortions. In this work, we present an \textit{ab-initio} investigation of the structural, electronic and hyperfine properties of the entire NaRTiO$_{4}$ series across the two disputed ground states, $Pbcm$ and $P\bar{4}2_1m$, and the high temperature $P4/nmm$ symmetries. Our results reveal an ionic-radius-dependent evolution from a tilt-dominated regime for small rare-earth ions to a distortion-dominated regime for larger cations, leading to an asymptotic regime in which the high-temperature phase becomes increasingly competitive with the ground-state structures as the ionic radius increases. In parallel, the electronic band gap follows a systematic evolution across the series, reflecting the underlying structural changes and the increasing dominance of octahedral distortions at larger ionic radii. The Electric Field Gradient (EFG) tensor reveals that, in the large-radius limit, all symmetries tend locally towards a similar  environment. Away from this limit, the EFG tensor for different symmetries progressively diverges, providing a sensitive probe for phase transitions and revealing symmetry-specific fingerprints, particularly for the rare-earth and Ti sites. By establishing these EFG signatures, this work provides a roadmap for experimental techniques, such as Nuclear Magnetic Resonance (NMR) and Perturbed Angular Correlation (PAC), to resolve the ground-state symmetry of these structures.

%Old
%\noindent The $n = 1$ Ruddlesden-Popper titanates, NaRTiO$_{4}$ (R = rare-earth), exhibit a structural behaviour where non-centrosymmetry is driven by cooperative oxygen octahedral rotations (OORs) rather than conventional second-order Jahn-Teller distortions. In this work, we present an \textit{ab-initio} investigation of the structural, electronic and hyperfine properties of the entire NaRTiO$_{4}$ series across the two disputed ground states, Pbcm and P421m, and the high temperature P4/nmm symmetries. \textbf{Our results show a radius-dependent behaviour where the small ion tilting dominated regime transitions into a distortion favoring one for larger cations, leading to a convergence where the high temperature phase becomes asymptotically as favorable as the ground state candidates with increasing ionic radius}. We demonstrate that the Electric Field Gradients (EFG) tensor serves as a sensitive probe of these transitions, identifying symmetry specific fingerprints that may help pinpoint the true ground state through experimental means. Notably, Yttrium emerges as a structural outlier, though the origins of these behaviour remain elusive. By establishing these EFG signatures, this work aims to provide a preemptive roadmap for techniques such as NMR and PAC to definitively resolve the ground-state of these structures.

\end{abstract}

%\pacs{71.45.Gm,36.40.Cg,61.50.Ks,61.50.Ks,77.80.Jk,64.60.ah}
%\keywords{Octahedral Rotations, Naturally Layered Perovskites, Negative Thermal Expansion}

% Print title, authors and abstract
\maketitle

% ---------------- Table of contents (optional) ----------------
% For drafts, TOC is useful. Remove it for final submissions.
%\tableofcontents

% ---------------- Header/footer marks ----------------

\markboth{}{}

% ---------------- Section numbering style in the main paper ----------------
% The paper uses Roman numerals for top-level sections (I, II, ...)
% and a custom subsection label (``I - A''). Change if you prefer standard numbering.
\renewcommand{\thesection}{\Roman{section}}
\renewcommand{\thesubsection}{\Alph{subsection}  }

% -------------------- MAIN BODY --------------------
% Each logical chunk of the paper is placed in a separate file under ./sections/.
% This keeps the project modular: each collaborator can edit their section file.

\section{Introduction}
\label{sec:intro}

The structural and electronic versatility of Naturally Layered Perovskites (NLPs) has positioned them as a cornerstone for the development of next-generation functional materials, enabling applications ranging from catalysis to electronics\cite{tayari2025comprehensive,tang2023photoexcited,uchino2015glory,park2020advances,bousquet2016non,fiebig2016evolution,rocha20202}. Among these, $n$=1 %\textit{
Ruddlesden-Popper (RP) titanates of the form ARTiO$_{4}$ (A = H, Li, Na, K, Ag; R = rare-earth) have emerged as a unique platform for studying unconventional symmetry breaking. Unlike traditional perovskites where non-centrosymmetry is often driven by Second-Order Jahn-Teller (SOJT)~\cite{bersuker2013pseudo} active cations, these titanates have been shown to exhibit acentric behavior arising from cooperative Oxygen Octahedra Rotations (OORs), providing a rare pathway to stabilize non-centrosymmetric structures, which account for only a small fraction of known oxides \cite{karen2002inorganic}. Furthermore, a recent study has shown that A-site cation substitution, %a substitution of the A-site cation, 
particularly for the AgRTiO$_{4}$, stabilizes the usually competing coexistence of both OORs and Oxygen Octahedra Deformations (OODs)\cite{yoshida2022interplay}, resulting in unique physical phenomena such as Biaxial Negative Thermal Expansion (BNTE), with piezoelectricity arising for the Li and Na analogues\cite{gupta2016improper,akamatsu2014inversion}, highlighting the relevance of the underlying structural mechanisms in these compounds.

Historically, the NaRTiO$_{4}$ series (R = Y, Ln), at low temperatures,  was assigned to the centrosymmetric $Pbcm$  space-group (\#57) \cite{toda1996crystal,balachandran2014crystal}. However, recent investigations by Akamatsu \textit{et al.}\cite{akamatsu2014inversion} challenged this paradigm, suggesting that several variants undergo a structural phase transition to an acentric, non-polar tetragonal  $P\bar{4}2_{1}m$ (\#113) ground state. This transition, from the  $P4/nmm$ centric stage, which possesses no rotations, is governed by ($a^{−}b^{0}c^{0}$)/($b^{0}a^{−}c^{0}$) type rotations that effectively remove inversion centers without inducing net polarization. The stability of this $P\bar{4}2_{1}m$ symmetry seems to be intimately linked to the ionic radius of the R-site cation, since Akamatsu group, through the use of Second Harmonic Generation (SHG), measured the phase transition temperatures of these compounds and showed that the critical temperature decreases with an increase in ionic radius.

Despite the progress made through long-range diffraction techniques, the NaRTiO$_{4}$ family often exhibits subtle local distortions that can be overlooked by conventional X-ray diffraction (XRD) and Raman Spectroscopy. To address these nuances, this work employs an approach where \textit{ab-initio} calculations allow one to model the hyperfine structure at each site. By systematically substituting lanthanides at the R-site, we aim to map the structural and electronic tendencies of these oxides, with a specific focus on the Electric Field Gradient (EFG) tensor.

The EFG is a highly sensitive probe of the local charge distribution and the immediate chemical environment of the nuclear site. By establishing a theoretical baseline for EFG behavior across the lanthanide series, this work aims to provide a critical reference framework for interpreting experimental hyperfine measurement results, such as Nuclear Magnetic Resonance (NMR), Mössbauer and Perturbed Angular Correlation (PAC) spectroscopies, with the use of established probes such as $^{23}$Na\cite{chubak2023quadrupolar}, $^{44}$Ti\cite{catchen1991investigating,budzinsky2017use} or $^{172}$Lu \cite{jancso2017tdpac}, that could match each atomic site in NaRTiO$_{4}$. Ultimately, this approach seeks to provide a path towards an unambiguous confirmation of the $P\bar{4}2_{1}m$  ground state and elucidate the local disturbances that define the functional potential of these layered titanates.

\section{Methodology}
\label{sec:methodology}
% Overview of computational and theoretical approach.

\subsection{Computational Details}
\label{sec:2-Methodology-DFT}
% List codes used, pseudopotentials, convergence parameters, k-point grids,
% energy cutoffs, smearing method, and any special settings.
% Example: "All DFT calculations were performed using X code with PBE functional..."
% !TEX root = ../main.tex
%All first-principles calculations were performed within the framework of density functional theory (DFT) \cite{hohenbergInhomogeneousElectronGas1964,kohnSelfConsistentEquationsIncluding1965} using the \textit{Quantum ESPRESSO} software suite \cite{giannozziQUANTUMESPRESSOModular2009,giannozziAdvancedCapabilitiesMaterials2017}. The exchange-correlation functional was described via the Perdew-Burke-Ernzerhof (PBE) generalized gradient approximation (GGA) \cite{perdewGeneralizedGradientApproximation1996}, while electron-ion interactions were modeled using projector-augmented wave (PAW) pseudopotentials \cite{blochlProjectorAugmentedwaveMethod1994}. These pseudopotentials demonstrate particular efficiency for electric field gradient (EFG) calculations \cite{petrilliElectricfieldgradientCalculationsUsing1998}, a critical consideration for this study.  

The calculations were performed using the \textit{Quantum ESPRESSO} software suite \cite{giannozziQUANTUMESPRESSOModular2009,giannozziAdvancedCapabilitiesMaterials2017}. The electronic interactions
were described within the framework of density functional theory (DFT) \cite{hohenbergInhomogeneousElectronGas1964,kohnSelfConsistentEquationsIncluding1965},  considering the Perdew-Burke-Ernzerhof (PBE) GGA functional \cite{perdewGeneralizedGradientApproximation1996}
to describe the exchange-correlation potential, while the electron-ion interactions were treated by the the projector-augmented wave (PAW) pseudopotentials \cite{blochlProjectorAugmentedwaveMethod1994}. Given the study's focus, PAW pseudopotentials were selected for their high performance and reliability in calculating electric field gradients (EFG) \cite{petrilliElectricfieldgradientCalculationsUsing1998}.

The valence electronic configurations for the pseudopotentials were specified as follows: \ce{Na}: $2s^{2}2p^{6}3s^{1}$; \ce{Ti}: $3s^{2}3p^{6}3d{^2}4s{^2}$; \ce{O}: $2s^{2}2p^{4}$; \ce{Y}: $4s^{2}4p^{6}4d{^1}5s{^2}$. For the lanthanide series, the 4$f$ electrons were treated as %being 
frozen in the core, following the established valence electronic configurations for the pseudopotential from Dal Corso \cite{dalcorsoPseudopotentialsPeriodicTable2014}:  $5s^{2}5p^{6}5d^{1}6s^{2}$. An exception to the rule was made for 
La, where the 4$f^{0}$ electrons were treated both:  frozen in the core, La$_{\text{core}}$ and  as semi-core states, La$_{\text{val}}$, with valence $5s^{2}5p^{6}4f^{0}5d^{1}6s^{2}$.

The initial structural models for NaRTiO$_{4}$ (R = Y, Ln) were based on experimental Crystallographic Information Files (CIF) obtained from Akamatsu \textit{et al.} \cite{akamatsu2014inversion}, for $P\bar{4}2_{1}m$ and $P4/nmm$ structures,  and from Toda \textit{et al.} \cite{toda1996crystal}, for the $Pbcm$ phase. For lanthanides lacking previously reported structures, initial models were constructed by substituting the R-site in the CIF of the lanthanide with closest ionic radius, according to values of the \textit{Database for Ionic Radii}\cite{shannon1976revised}. All structures were then screened through a SCF calculation, with electronic convergence was strictly enforced with a threshold of $1.0 \times 10^{-9}$~Ry and a mixing beta of 0.7, before being fully optimized via variable-cell relaxation (VCR) to determine the equilibrium lattice parameters and atomic positions for each specific R-site ion, with convergence achieved when residual forces fell below 0.01 Ry/au ($\approx$ 0.03 eV \AA$^{-1}$).

%With a fully relaxed structure, a second SCF calculation was performed followed by a non-self-consistent field (NSCF) run.
The plane-wave kinetic energy cutoffs of $E_{\text{cut}}^{\psi} = 80$ Ry for the wavefunctions and $E_{\text{cut}}^{\rho} = 320$ Ry for the charge density were employed. Brillouin zone integration was performed using a $(8 \times 8 \times 4)$ Monkhorst-Pack\cite{monkhorst1976special} $k$-point mesh for the 28-atom cells (adapted for the 14-atom $P4/nmm$ structures).% while a denser $12 \times 12 \times 12$ grid with 132 bands was used for the NSCF stage. 

 Finally, the $V_{ij}$ electric field gradient (EFG) tensor components  were evaluated at all symmetry-inequivalent sites using the gauge-including projector-augmented wave (GIPAW) method \cite{charpentier2011paw}, along with the calculation of the electron localization function (ELF) and projected density of states (PDOS).

\subsection{Hyperfine Interactions and Electric Field Gradient}
\label{sec:2-Methodology-EFG}

Hyperfine interactions originate from the coupling between nuclear moments and extranuclear electromagnetic fields, resulting in the lifting of degeneracy in nuclear energy levels. In non-magnetic systems, the interaction is dominated by the electric quadrupole term, which arises from the coupling of the nuclear quadrupole moment to the local electric field gradient (EFG).

For a nucleus with a charge distribution $\rho_N(\mathbf{r})$, the electric quadrupole moment tensor $Q_{ij}$ measures the deviation from spherical symmetry and is defined as \cite{schatzNuclearCondensedMatter1996}:
\begin{equation}
    Q_{ij} = \frac{1}{e} \int \rho_{N}(\mathbf{r}') \left( 3x'_i x'_j - (r')^2 \delta_{ij} \right) d^3\mathbf{r}'
\end{equation}
where $e$ is the elementary charge. In a quantum mechanical framework, this tensor can be expressed via the nuclear spin operators $I_i$:
\begin{equation}
    Q_{ij} = \frac{eQ}{I(2I-1)} \left[ \frac{3}{2}(I_i I_j + I_j I_i) - \delta_{ij} I(I+1) \right]
\end{equation}
The scalar quantity $eQ$ is the spectroscopic quadrupole moment, representing the expectation value of the operator in the state $|I, M=I\rangle$. Notably, symmetry arguments governed by the Wigner-Eckart\cite{schatzNuclearCondensedMatter1996,suhonen2007nucleons} theorem imply that $Q$ vanishes for nuclei with spin $I < 1$.

The interaction between the quadrupole moment and the EFG tensor $V_{ij}$ (defined as the second spatial derivative of the electrostatic potential $\Phi$) is described by the Hamiltonian:
\begin{equation}
    \mathcal{H}_Q = \frac{e}{6} \sum_{i,j} V_{ij} Q_{ij}
\end{equation}
where $V_{ij}$ represents the EFG tensor components at the nuclear site \cite{schatzNuclearCondensedMatter1996}. The EFG tensor is symmetric and traceless in a Laplacian potential. In its principal axis system, it is characterized by the principal component $V_{zz}$ and the dimensionless asymmetry parameter $\eta$:
\begin{equation}
    \eta = \frac{V_{xx} - V_{yy}}{V_{zz}}, \quad (|V_{zz}| \geq |V_{yy}| \geq |V_{xx}|)
\label{etadefinition}
\end{equation}
where $0 \leq \eta \leq 1$. While $V_{zz}$ reflects the strength of the gradient due to local charge anisotropy, $\eta$ quantifies the deviation from axial symmetry ($\eta = 0$).

Substituting these parameters, the Hamiltonian becomes:
\begin{equation}
    \mathcal{H}_{Q} = \frac{eQV_{zz}}{4I(2I-1)} \left[ 3I_{z}^2 - I(I+1) + \eta(I_{x}^{2} - I_{y}^{2}) \right]
\end{equation}
For an axially symmetric system ($\eta = 0$), the transition frequencies between sublevels $|M\rangle$ and $|M'\rangle$ are integer multiples of the quadrupole frequency $\omega_Q = \frac{eQV_{zz}}{4I(2I-1)\hbar}$. The fundamental observable frequency is typically reported as $\nu_Q = eQV_{zz}/h$, which serves as a direct probe of the local electronic environment.

%The PAC experiments after performing ion-beam $^\text{111m}$Cd implantation were conducted at CERN-ISOLDE\cite{schellPerturbedAngularCorrelations2017} in a polycrystalline sample that was synthesized following the procedure detailed in Ref.\cite{zhuComplexStructuralPhase2020}. 
%As there is already Cd-based DJ perovskite structures\cite{atriSynergisticInfluence02020}, and despite the $^\text{111m}$Cd probe has an oxidation state ($+2$) that is not present in \ce{CsNdNb2O7}, and its key applicability in previous works\cite{rocha-rodriguesCa2Mn2020,rocha-rodriguesCa3Mn2O7StructuralPath2020,rocha-rodriguesProbing$mathrmCa_3mathrmTi_2mathrmO_7$Crystal2024} of the same scope make it an excellent probe for this analysis. 

%Via the measurement of perturbation functions, the fundamental quadrupole frequency $\omega_0$ can be calculated, in order to calculate the principal component $V_{zz}$, a nucleus electric quadrupole moment ($Q$) of 0.664 b\cite{haasFreeMoleculeStudies2021} was used in this work.

% \subsection{Enthalpy of Formation}
% \label{sec:2-Methodology-Hf}
% Provide formulae and references for how formation enthalpies were computed.
% \input{./sections/2-methods_enthalpy.tex}

\section{Results}
\label{sec:results}
% Break results into clear, self-contained subsections. Each subsection file should
% produce figures and tables with labels for cross-referencing from the text.

\subsection{Structural Properties}

% Energia e estabilidade de cada fase, comparar com SHG Akamatsu, notar a tendencia raio ionico e temperatura de transicao
\subsubsection{Energetic Stability}
\label{sec:resultsESTab}
\begin{figure}[b]
	%\centering
    \includegraphics[width=1\columnwidth]{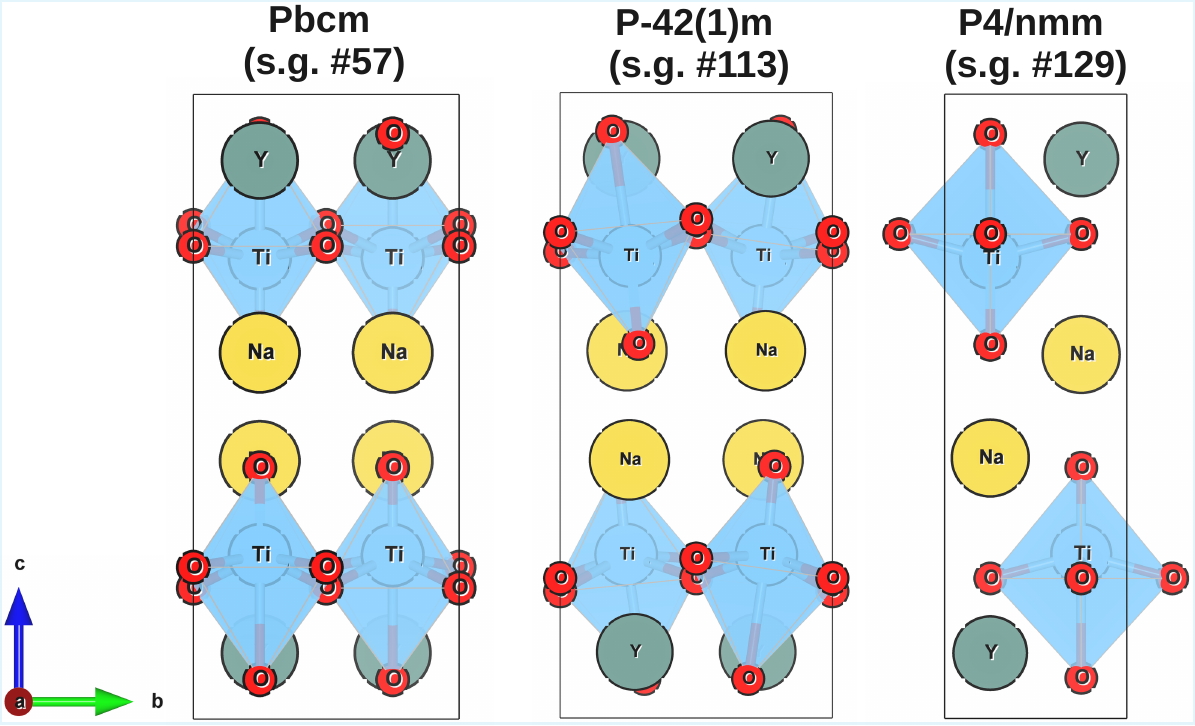}
	\caption{\justifying 
		Representation of the unit cell of all three phases adopted: the orthorhombic $Pbcm$ (s.g. \#57); acentric $P\bar{4}2_{1}m$  (s.g. \# 113), and the tetragonal $P4/nmm$ (s.g. \#129) %high temperature phase. 
        In this illustration, the R-site in NaRTiO$_{4}$ is occupied by the ytrium atom. Figures plotted with the help of VESTA software \cite{momma2011vesta}.
        }
	\label{fig:1-symmetries}
\end{figure}

\begin{figure*}[htb]
  \centering
  % Subfigure (a): Non-Magnetic (NM) Reference Structure
  \begin{subfigure}[b]{0.45\textwidth}
    \centering
     \includegraphics[width=\linewidth]{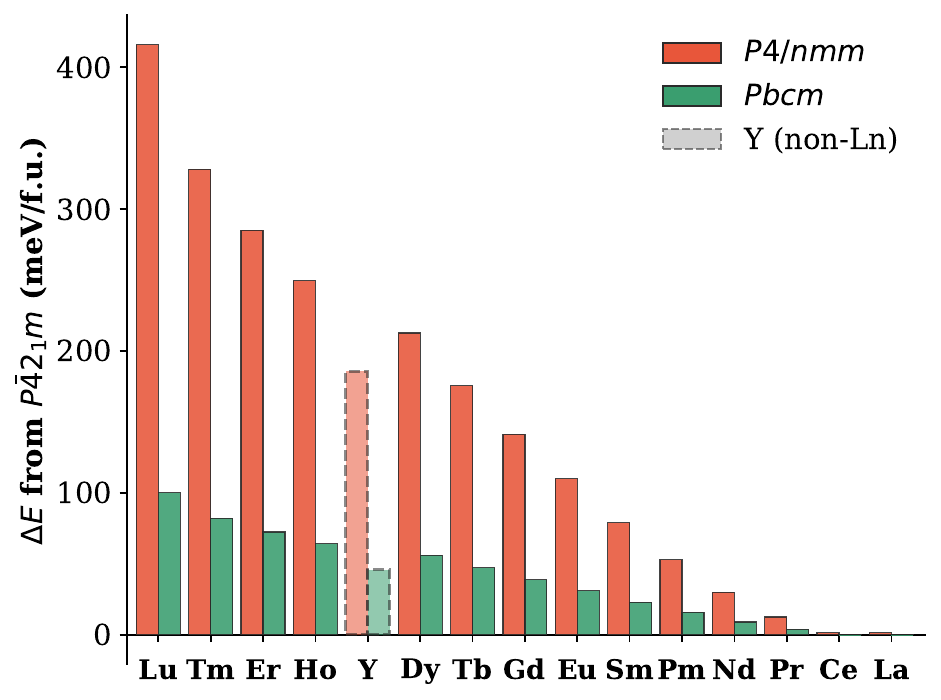}
    \caption{DFT-calculated ground-state energy of NaRTiO$_{4}$ compounds, ordered by increasing ionic radius, for the $P4/nmm$ and $Pbcm$ phases. These values are taken relative to the reference energy of the acentric $P\bar{4}2_{1}m$  phase.}
    \label{fig:Stability_vs_Raio}
  \end{subfigure}
  \hfill
  % Subfigure (b): 
  \begin{subfigure}[b]{0.45\textwidth}
    \centering
    \includegraphics[width=\linewidth]{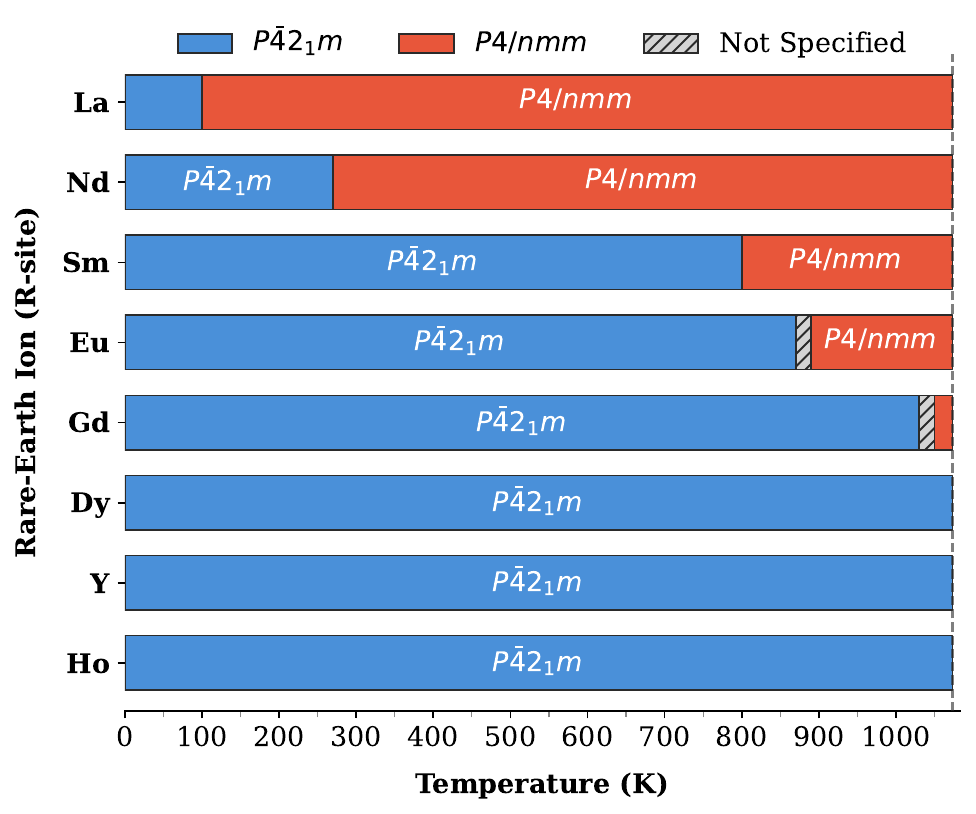}
    \caption{Reported phases and transition temperatures of the NaRTiO$_{4}$ family of structures, measured through Second Harmonic Generation (SHG) up to $T = 1073 $K\cite{akamatsu2014inversion}.}
    \label{fig:SHG_PhaseTransition}
  \end{subfigure}

  \caption{
   Phase stability and structural evolution of the NaRTiO$_{4}$ family. (a) Relative DFT-calculated ground-state energies for the $P4/nmm$ and $Pbcm$ phases as a function of the rare-earth ionic radius, benchmarked against the acentric $P\bar{4}2_1m$ phase. Yttrium is shown with dashed lines in order to stand out further from the Lanthanide trend. (b) Experimental phase transition map determined via Second Harmonic Generation (SHG) measurements up to 1073~K.
  }
  \label{fig:two‐structures_consolidated}
\end{figure*}

As mentioned in Section~\ref{sec:intro}, this work focuses on the structural and electronic trends across the NaRTiO$_{4}$ series, with the primary objective to identify a pathway through which hyperfine measurement techniques could resolve the definitive ground state of these compounds, more specifically by identifying distinct signatures across the rare-earth series.

To achieve a reliable analysis of the EFGs, it is first necessary to establish a robust structural model. This requires a systematic evaluation of structural stability in comparison with existing literature. We begin by examining the total energy for the three candidate symmetries: $P4/nmm$, $Pbcm$, $P\bar{4}2_{1}m$,  unit cells shown in  Figure~\ref{fig:1-symmetries}. %, illustrated in Figure~\ref{fig:Stability_vs_Raio}.

By using the acentric $P\bar{4}2_{1}m$ phase as the total energy reference, Figure~\ref{fig:Stability_vs_Raio} illustrates the energy difference among the three space groups. A clear trend emerges: as the ionic radius ($r_{i}$) increases, the relative energies of the $P4/nmm$ and $Pbcm$ phases decrease significantly (in meV/f.u.). Furthermore, for the larger cations Ce and La, these energy differences are nearly zero, suggesting that the phases become nearly equally energetically favorable. This observation corroborates the findings of Akamatsu \textit{et al.} \cite{akamatsu2014inversion}, who reported that for La, both the $Pbcm$ and $P\bar{4}2_{1}m$  structure calculations tended to \textit{relax} toward the higher-symmetry $P4/nmm$ configuration. Notably, while we were unable to sucessfully converge the La$_{\text{core}}$ configuration, the La$_{\text{val}}$ results align with the trends observed for the other lanthanides where the 4$f$ shell was considered frozen in the core, as we will see in the coming sections. This suggests that the 4$f$ orbitals do not exert a significant influence on the structural stability nor the electrical field gradients in these structures.

This calculated trend is then further supported by previous experimental evidence. SHG measurements\cite{akamatsu2014inversion} on several rare-earth entries of this family of materials, shown in Figure~\ref{fig:SHG_PhaseTransition}, show that as an increase in $r_{i}$ decreases, the $P4/nmm$ phase becomes significantly less energetically stable relative to $P\bar{4}2_{1}m$ , which in turn leads to an increase in the critical temperature for the acentric to centric phase transition. As shown above, SHG signals, which were taken up to 1073~K, suggest that while Y, Dy, and Ho remain acentric within the measured range, they are expected to transition to the $P4/nmm$ phase at sufficiently high temperatures, continuing the trend of the other rare-earths.

Furthermore, although not being a lanthanide, Yttrium possesses a similar charge state (often +3 with IX coordination in NaRTiO$_{4}$) as the rest of the Ln series, with ionic radius in-between Dy and Ho. However, as seen in Figure~\ref{fig:Stability_vs_Raio}, it breaks the evolution of phase stability with ionic radius established by the rest of the lanthanides, aligning more closely with Gd and Tb than its immediate $r_{i}$ neighbors, suggesting that factors beyond simple charge and size dependencies are at play. 
.

% lattice parameters, comparar tambem aos resultados do akamatsu notando que esta igual, da rigor ao modelo, comentar a trend que se observa em funcao do raio ionico e entre as fases, reforçar aqui a postular a diferença estrutural do Yttrio.
\subsubsection{Equilibrium Lattice Parameters}
\label{sec:resultsLatParams}
\begin{figure}[h]
	\centering
	\includegraphics[width=1\columnwidth]{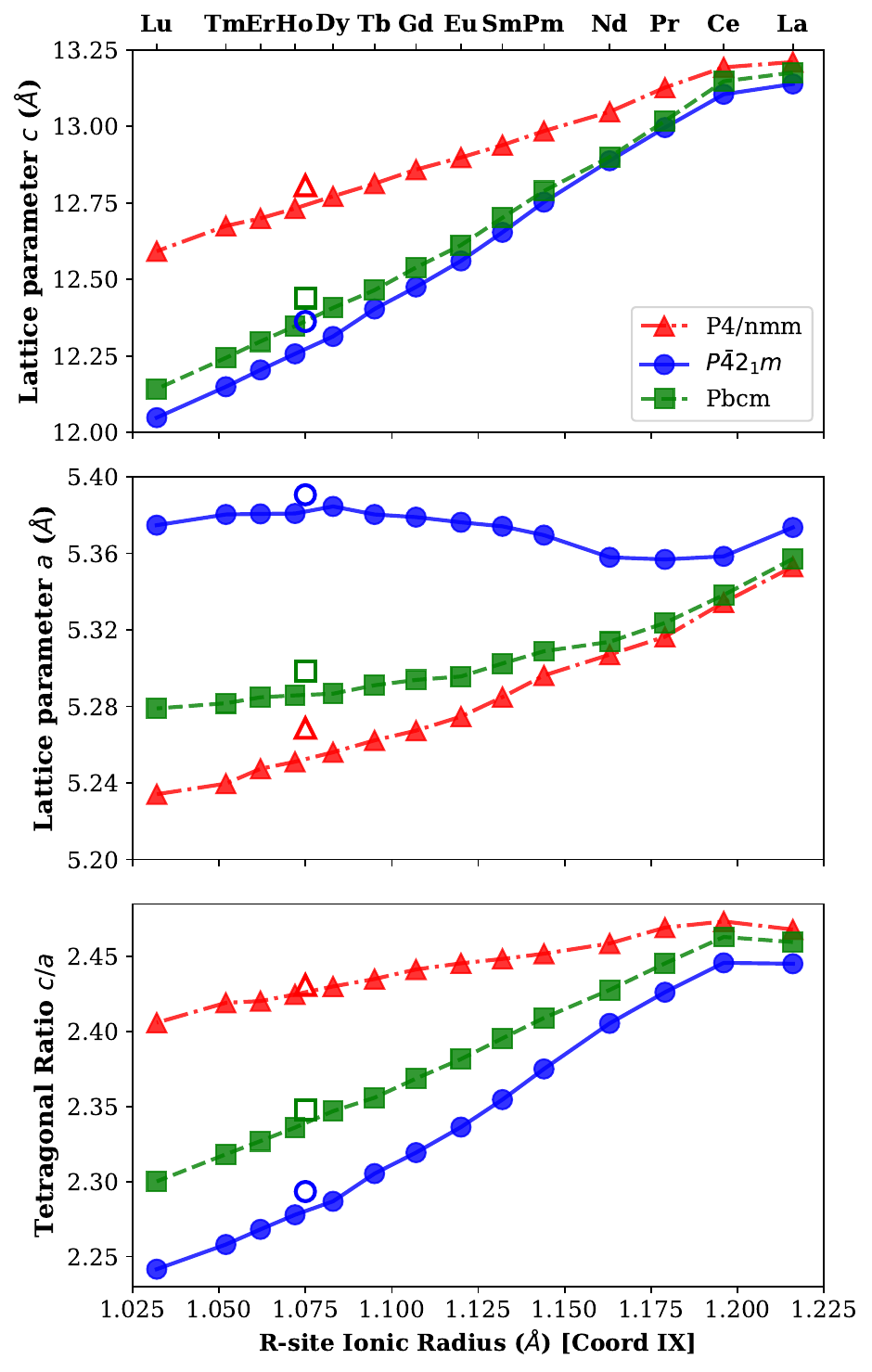}
	\caption{%
		DFT-calculated cell parameters of the NaRTiO$_{4}$ (R = rare-earth) series, as a function of ionic radii ($r_{i}$). From top to bottom: Long-axis (c) ; Short-axis (a) and Tetragonal Ratio (c/a). All three symmetries are highlighted in distinct colors: P4/nmm (red); P$\bar{4}$2$_{1}$m (blue); Pbcm (green).  Yttrium data points are represented by hollow markers.
        }
	\label{fig:3-structural_params}
\end{figure}

In order to consolidate the proposed theoretical model, and to verify their consistency before moving onwards to a more localized electronic analysis, we performed a systematic evaluation of the equilibrium (post-VCR) cell parameters for the NaRTiO$_{4}$ series. By examining these parameters as a function of $r_{i}$ we aim to establish a structural baseline that can be directly compared against existing literature. The results of this structural optimization are shown in Figure~\ref{fig:3-structural_params}.

When compared to the data obtained by Akamatsu's group \cite{akamatsu2014inversion}, we see that the model demonstrates a high fidelity when compared to their experimental results, particularly for P4/nmm. For the remaining two symmetries, the calculated values are somewhat overestimated, though they remain within 2\% of experiment in the most critical cases (typically associated with the long-axis (c)). This divergence however, is likely a functional artifact. While Akamatsu's group utilized \textit{VASP}\cite{hafner1997vienna} with the PBEsol\cite{perdew2007generalized} functional, which is specifically aimed at solid-state lattice constants, our implementation employs the PBE functional within \textit{Quantum ESPRESSO}. Even so, the slight overestimation of unit cell volumes is consistent with the well known tendency of PBE to overestimate bond lengths and cell volumes, and as such the close agreement with both experimental and DFT-calculated values confirms that our model is sufficiently robust to move onto a more detailed analysis of these systems.

Looking over at Figure~\ref{fig:3-structural_params}, a clear trend is observed in the lattice evolution. Both the short ($a$) and long ($c$) axes expand as the rare-earth $r_{i}$ increase. Notably, the $c$-axis exhibits a much steeper growth relative to the short axis across the series. Furthermore, if we observe each symmetry separately, we can see that P$\bar{4}$2$_{1}$m has the largest short-axis, $a$, of the three candidate phases, which seems to remain very stable across the entire series, in contrast to the previously mentioned increase of the remaining two symmetries. By comparison, P4/nmm displays the most contracted short axis but the most elongated $c$-axis, with Pbcm remaining intermediate on both parameters. 

Crucially, these competing trends, where a longer $c$-axis seems to be compensated by a shorter $a$-axis, result in a nearly uniform cell volume across all three symmetries. Because the cell volume scales linearly with $r_{i}$ and is nearly identical across all phases, it appears to be a poor descriptor for probing structural phase transitions in the NaRTiO$_{4}$ series. 

In contrast to the volume, we observe that as the ionic radius decreases, the gap in the tetragonal ratio, $c/a$, between the symmetries widens significantly, with P4/nmm being the most elongated of the 3 phases, and P$\bar{4}$2$_{1}$m the least. This suggests that the tetragonal ratio could be used as a probing mechanism for identifying the onset of phase transitions in the series, particularly for smaller rare-earth ions where these distinctions are more prominent. Conversely, in the large-ion limit (i.e. Ce; La), we see that the $c/a$ ratios remain nearly constant across all three symmetries, which corroborates with our previous analysis of the energetic stability, indicating that the competitive advantage of any single phase is reduced for larger ions, leading to a similar behaviour between the three candidates. 

Finally, it is worth noting that similar to what was discussed in Section~\ref{sec:resultsESTab}, NaYTiO$_{4}$ represents an outlier in these trends, displaying a localized expansion both in the short and long axes, as well as a $c/a$ ratio and cell volume above those of its direct neighbours, with the cell volume being considerably larger. This suggests that the structural chemistry of the Y-site involves coordination effects or orbital bonding that deviate from the seemingly linear evolution observed in the lanthanide series, a factor whose implications may stem from its localized electronic properties.

% Falar do Ti-O Displacement e da Distorçao e como esta varia em funcao de raio ionico e fase etc
\subsubsection{Octahedral Distortion and Ti Displacement}
\label{sec:resultsOctDistortion}
Having established that our structural model agrees with literature, we now examine the local coordination environment of the Ti-site. A rigorous characterization of the TiO$_{6}$ octahedra serves as a useful prerequisite for the electronic structure analysis, as the local symmetry and Ti-displacement are known to be drivers of electronic properties in perovksite structures \cite{shirsath2022interface}.

\begin{figure}[h]
	\centering
	\includegraphics[width=1\columnwidth]{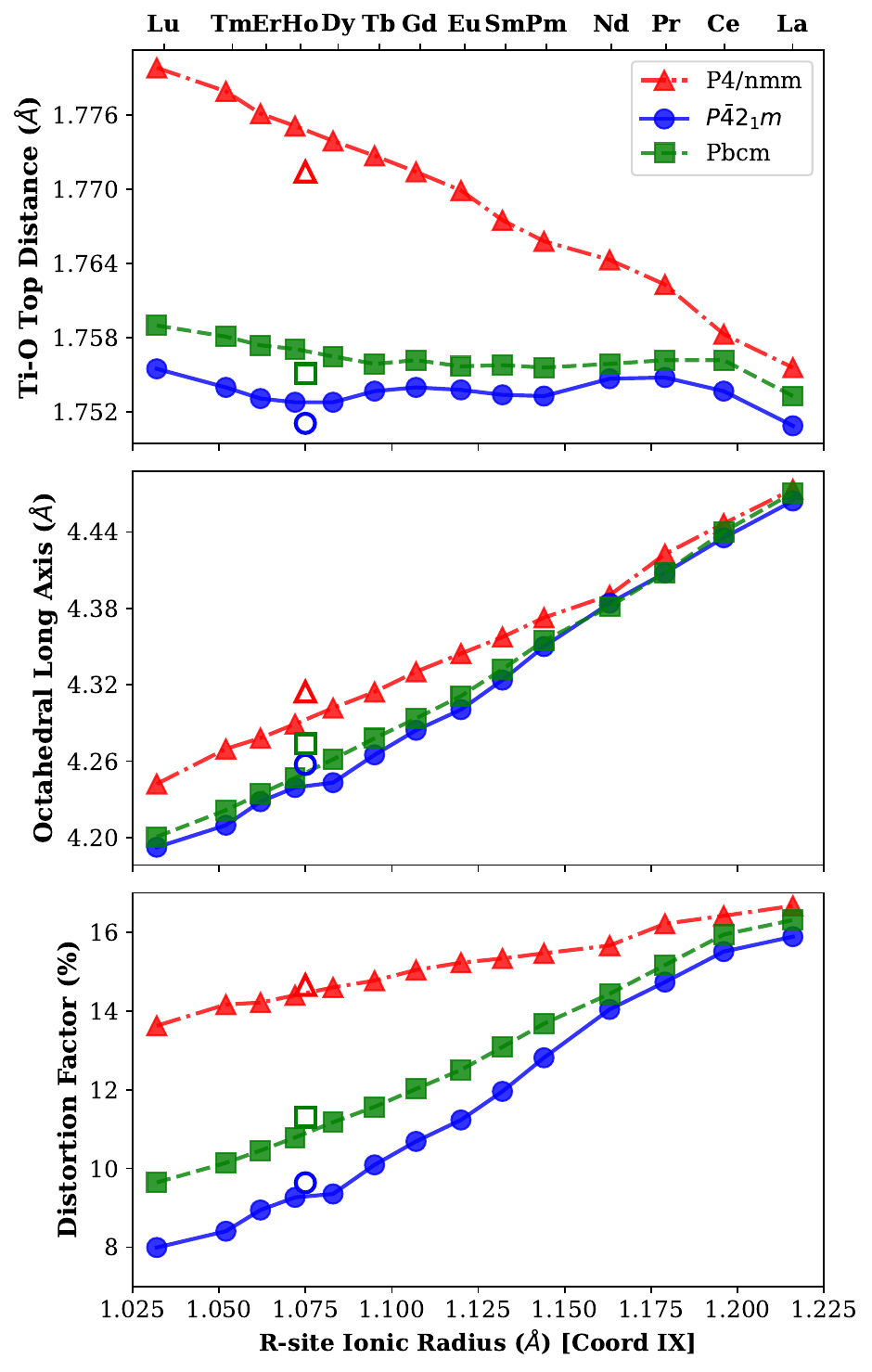}
	\caption{%
		Evolution of the local TiO$_{6}$ environment in the NaRTiO$_{4}$ series, as a function of the ionic radius. (Top) Displacement of the Ti ion toward the apical oxygen, Ti--O$_{\text{top}}$, where O$_{\text{top}}$ is defined as the oxygen neighbor closest to the Na layer. (Middle) Total vertical height ($L_v$) of the TiO$_{6}$ octahedron. (Bottom) Octahedral deformation factor ($\sigma$). All three symmetries are highlighted in distinct colors: P4/nmm (red); P$\bar{4}$2$_{1}$m (blue); Pbcm (green).  Yttrium data points are represented by hollow markers.
        }
	\label{fig:3-Ti_distortion}
\end{figure}

To quantify the evolution of the local environment, we assess three key parameters as functions of the rare-earth sites' ionic radius, namely the Ti-site displacement towards the apical oxygen, the vertical height of the octahedron, and the octahedral deformation factor, $\sigma$\cite{yoshida2022interplay}, whose results are illustrated in Figure~\ref{fig:3-Ti_distortion}, and is defined as:
\begin{equation}
\sigma = \frac{L_v - L_h}{(L_v + L_h)/2} \times 100 (\%)
\end{equation}

where $L_{v}$ and $L_{h}$ are defined as the apical and equatorial dimensions of the TiO$_{6}$ octahedron, respectively. Under this definition, a positive $\sigma$ denotes an octahedron that is elongated along the $c$-axis, while a negative one reflects axial compression.

The first panel in Figure~\ref{fig:3-Ti_distortion} tracks the Ti--O$_{\text{top}}$ distance, which is defined as the bond length between the Ti cation within the oxygen octahedra and the apical oxygen atom closest to the Na layer (see Figure~\ref{fig:1-symmetries}). For Pbcm and P$\bar{4}$2$_{1}$m, we observe a subtle and nearly identical decline in this distance as the ionic radius increases, indicating a slight migration of the Ti ion toward the sodium layer. In contrast, the P4/nmm phase shows a linear decline with a significantly steeper slope, suggesting that for larger R-site ions the Ti cations in P4/nmm are displaced significantly further toward the Na layer and away from the rare-earth layer, compared to the remaining 2 symmetries, though this discrepancy diminishes as we reach the larger ions (i.e. Ce; La).

Complementing this displacement is the total vertical height of the TiO$_{6}$ octahedron. As expected, the vertical height of the octahedron increases with the increase in ionic radius, with P4/nmm consistently displaying the most elongated octahedra, followed by Pbcm and P$\bar{4}$2$_{1}$m, respectively.

Our results indicate that the evolution of $\sigma$ is similar that of the trend observed for the unit cell's tetragonal ratio ($c/a$). As the rare-earth ionic radius increases, the strain within the system that forces the TiO$_{6}$ to tilt is alleviated, allowing the TiO$_{6}$ octahedra to undergo progressive axial elongation. This stretching does not seem to be a merely passive response to the expanding lattice but also a primary driver of said behavior. In the P4/nmm phase, where this is most pronounced, the significant displacement of the Ti ion toward the apical oxygen, coupled with the steep rise in deformation factor,$\sigma$, points towards a mechanism by which the structure accommodates the larger ions by favoring a stretching of the octahedra rather than a tilting, thus leading to an approximation of the local environment of each phase as has been discussed for previously, pairing well with the fact that the P4/nmm does not allow for any OORs, only deformations (see Figure~\ref{fig:3-Results/TiltDistortion}).

\begin{figure*}[!htbp]
\centering
\includegraphics[width=\textwidth]{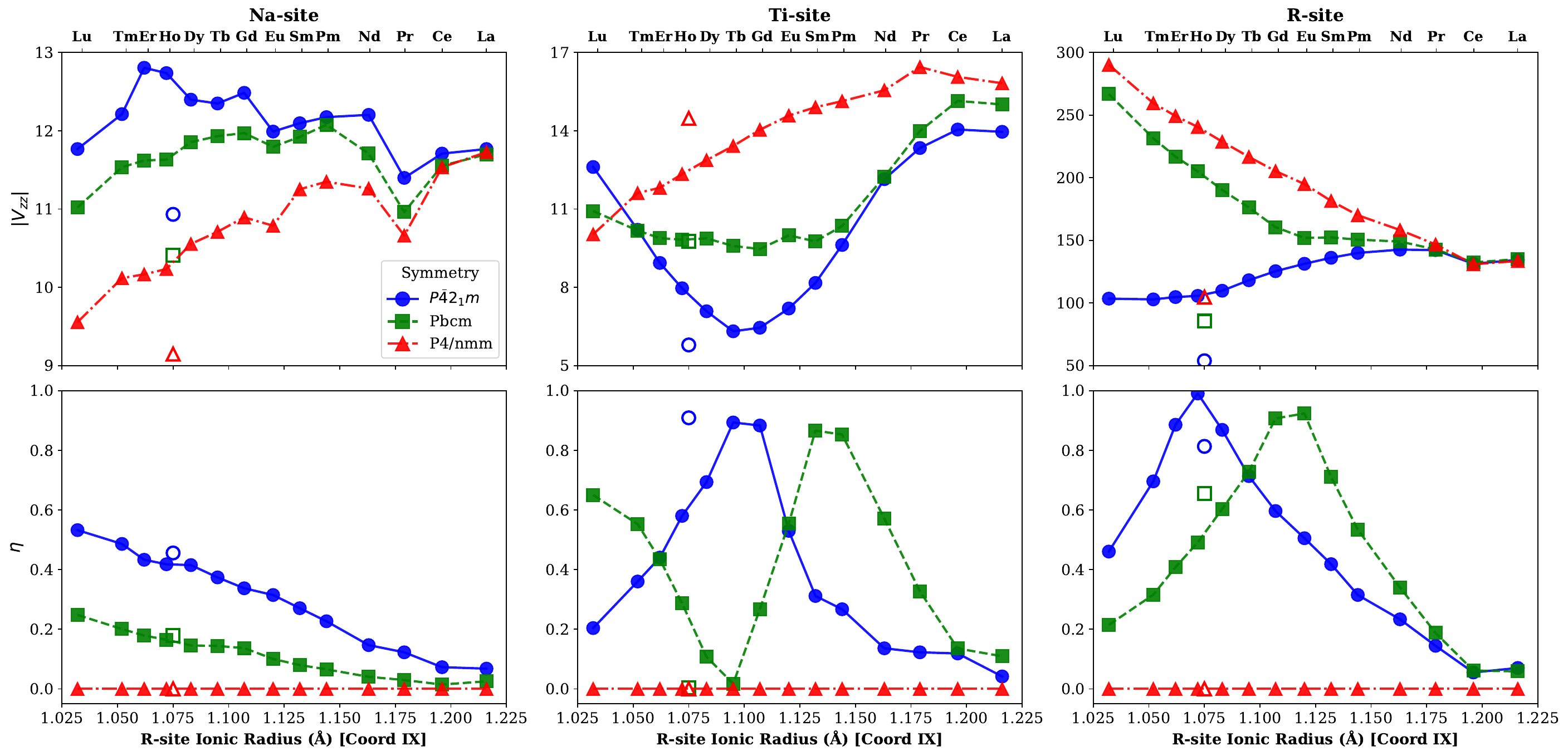}
\caption{Theoretical electric field gradient (EFG) parameters in NaRTiO$_{4}$ (R = rare-earth), as functions of ionic radius, $r_{i}$ for different atomic sites: (Left) Na, (Middle) Ti, and (Right) R. The upper panel shows the absolute principal component $|V_{zz}|$ (in $V  \AA^{-2}$), while the lower panel displays the asymmetry parameter $\eta$, both calculated for the three symmetries in study: P$\bar{4}$2$_{1}$m (blue), Pbcm (green), P4/nmm (red). Yttrium data points are represented by hollow markers.}
\label{fig:3-Results-DFT-EFGs/VZZ_ETA_RADIUS}
\end{figure*}

This is further supported by the outlier behaviour of Yttrium in the deformation factor plot, aligning with the previous hypothesis that the noted expansion of the $c$-axis is a direct manifestation of the strain in the octahedral layers. Because the unit cell must accommodate these elongated TiO$_{6}$ octahedra while maintaining the symmetry, the lattice is forced to stretch along the stacking direction.

%\begin{figure*}[tb]
%	\centering
%	\includegraphics[width=2\columnwidth]{figures/3-EFGsResults/VzzEta_IonicRadius_AllSites.pdf}
%	\caption{%
%		EFG Principal components of Nd atom for %studied magnetic orderings.
 %       }
%	\label{fig:3EFGs_IonicRadius}
%\end{figure*}

\subsection{Electronic Properties}
% Subdivide by property so readers can quickly find DOS, bandstructures, etc.
%\subsubsection{Partial Density of States}
%\label{sec:resultsPDOS}
%\input{sections/3-results_PDOS}

\subsubsection{Electric Field Gradients}
\label{sec:resultsEFG}
%\begin{figure*}[!htbp]
%\centering
%\includegraphics[width=\textwidth]%{figures/3-EFGsResults/VzzEta_IonicRadius_AllSites.pdf}
%\caption{Theoretical electric field gradient (EFG) parameters in NaRTiO$_{4}$ (R = rare-earth) for different atomic sites: (Left) Na, (Middle) Ti, and (Right) R. The upper panel shows the absolute principal component $|V_{zz}|$ (in $V  \AA^{-2}$), while the lower panel displays the asymmetry parameter $\eta$, both calculated for the three symmetries in study: P$\bar{4}$2$_{1}$m (blue), Pbcm (green), P4/nmm (red).}
%\label{fig:3-Results-DFT-EFGs/VZZ_ETA_RADIUS}
%\end{figure*}

The principal component $V_{zz}$ and asymmetry parameter $\eta$ of the electric field gradient (EFG) tensor were computed at Na, Ti and R (Y; Ln) sites across the three candidate symmetries, with comparative results visualized in Figure~\ref{fig:3-Results-DFT-EFGs/VZZ_ETA_RADIUS}.

At the Na and Ti sites, the magnitude of $V_{zz}$ remains relatively small across the entire series. For Na specifically, while P$\bar{4}$2$_{1}$m consistently yields slightly higher values, followed by Pbcm and P4/nmm, the differences are too subtle to provide any reliable experimental metric for distinguishing between the three proposed phases. As the ionic radius increases, the $V_{zz}$ values for all three symmetries converge, a recurring theme across all three atomic sites, reflecting the structural homogenization discussed in previous sections, where the energy and lattice parameters of the candidate phases become similarly favorable.

The Ti site exhibits more complex trends, with both P$\bar{4}$2$_{1}$m  and Pbcm following concave trends that peak at intermediate radii, namely Tb and Gd. While P4/nmm generally displays the largest $V_{zz}$ across the series, it steadily declines with decrease in $r_{i}$, eventually being surpassed by the other two symmetries at the small ion limit (Lu). 

Moreover, at the rare-earth (R) site a larger distinction can be observed, where both P4/nmm and Pbcm show magnitude values nearly three times as large as those measured for P$\bar{4}$2$_{1}$m for the smaller radii. However, this discrepancy is diminished as we move towards the larger ions, since Pbcm and P4/nmm follow a monotonic decrease with increasing radius, contrasted with P$\bar{4}$2$_{1}$m rising followed by a plateau starting around Eu and Sm. This distinct signature of P$\bar{4}$2$_{1}$m at the R-site suggest that a viable experimental pathway for distinguishing between the two previously reported ground states, through local probe techniques. Notably, Yttrium consistently acts as a significant outlier across all sites, deviating from the lanthanide trends even more sharply than what was observed in the structural analysis.

As for the asymmetry parameter, $\eta$, its values were identically zero at all sites for the P4/nmm phase, reflecting its high axial symmetry, expected of a high temperature phase. This in turn highlights a promising use of the EFG measurement as a probe for the transition onto P4/nmm in these systems, as the emergence of a non-zero $\eta$ during a decrease in temperature would signal a transition away from the P4/nmm phase to a lower-symmetry or a previously unreported intermediate phase. Regarding the Pbcm and P$\bar{4}$2$_{1}$m trends, the Na site remains unpromising as both symmetries follow similar trends, though a small gap persists for lower $r_{i}$ values. In constrast, the Ti and R sites for both symmetries exhibit peaks, where $\eta$ starts low at large ionic radii, close to the value seen for P4/nmm, and rises to a maximum at specific R-sites, subsequently declining. Since either symmetry peaks at a different R-site, mapping the position of this maximum across the series could further help confirm the true ground-state symmetry. These sudden changes in $\eta$ are linked to shifts in the principal components orientation, as we will discuss in Section \ref{sec:resultsVZZOrientation}.

When plotted against $\sigma$ instead of $r_{i}$ (see Figure~\ref{fig:3-Results/VzzEtaDistortion} in Appendix \ref{sec:EFG_DIST}), Yttrium is no longer an outlier in the $\eta$ trends of the lanthanides, for all three symmetries. This, however, is not seen for the $V_{zz}$ values, where Y still exhibits abnormally small values when compared to its neighbors, for a similar degree of distortion. This suggests that while the asymmetry of the charge distribution at the R-site is primarily a function of geometry and thus $\sigma$, the absolute magnitude of the gradient is also influenced by specific electronic traits unique to Yttrium, possibly related to the fact that Y is one of the fifth-period R elements and ultimately, chemically different from the Lanthanides. The R-O covalent bonds are attributed to Y's 4d, 5s orbitals, while they're governed by R 5d, 6s for the rest of the series.

On a final note, the $V_{zz}$($\sigma$) trends seem to converge from Sm onwards, very visible for Ti and which can be traced back to our discussion on octahedral deformation. As established before, larger rare-earth ions allow the TiO$_{6}$ units to elongate rather than be forced to tilt due to smaller R-site ionic radius' constraints, leading to a stretching of the long axis and favoring octahedral distortions over rotations. Once this distortion becomes the primary means of accommodating the cation, the local charge distribution around the nuclei becomes increasingly similar with the increase of $r_{i}$, leading to seemingly indistinguishable EFG signatures across the 3 symmetries for the larger ions.

\subsubsection{Principal Component Orientation}
\label{sec:resultsVZZOrientation}
\begin{figure}[t]
	\centering
	\includegraphics[width=1\columnwidth]{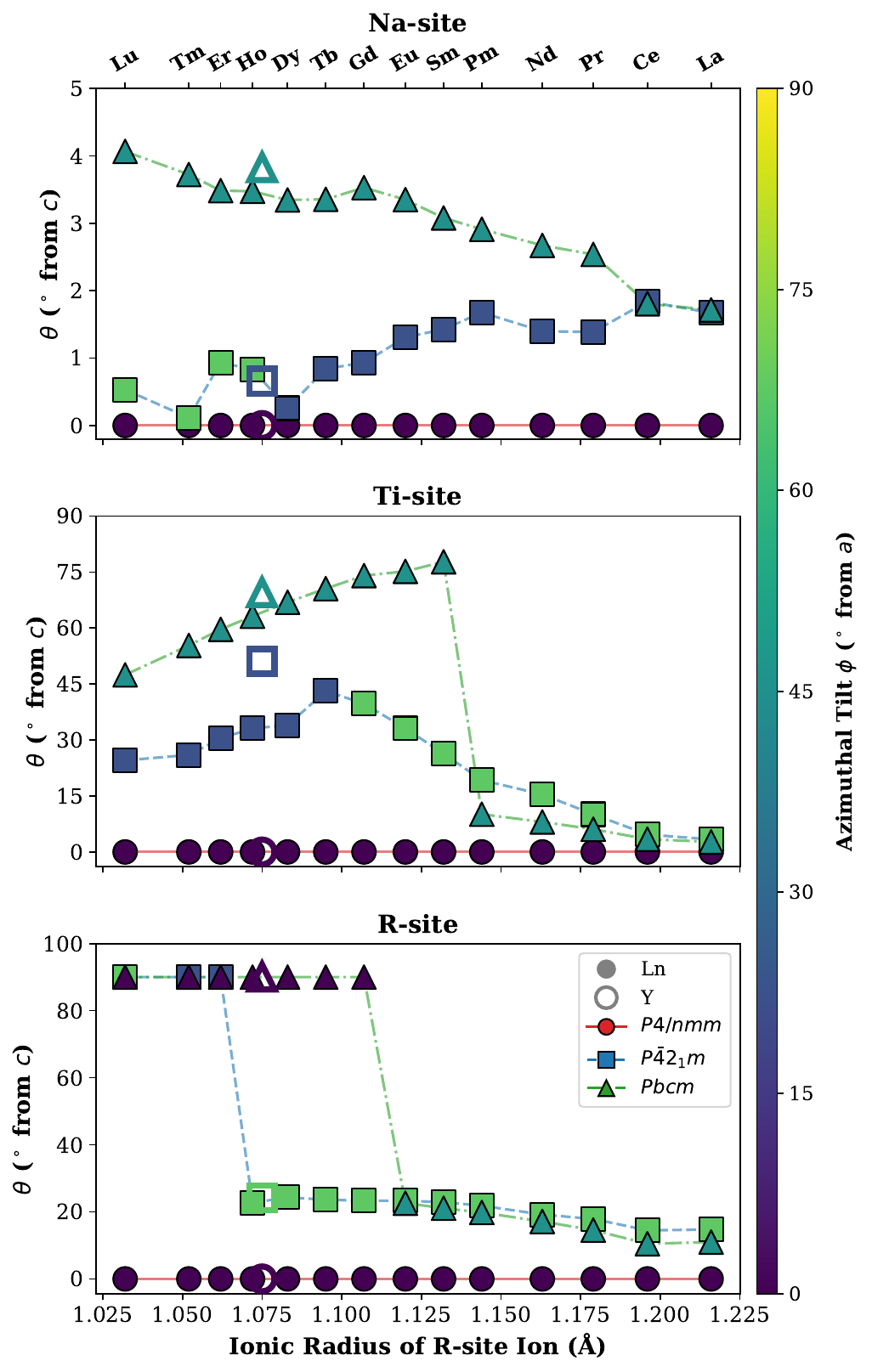}
	\caption{%
		Calculated values of the principal component, $V_{zz}$, orientation as a function of ionic radius, for the Na (top), Ti (Middle) and R (bottom) sites. The tilt angle ($\theta$) is shown by the y-axis while the colorbar refers to the azimuthal angle ($\phi$). All three symmetries are highlighted by trend lines with distinct colors: P4/nmm (red); P$\bar{4}$2$_{1}$m (blue); Pbcm (green). Yttrium data points are represented by hollow markers.
        }
	\label{fig:3-results-VzzOrientation}
\end{figure}

While the magnitude of the principal component and $\eta$ might help us quantify the local asymmetry, the spatial orientation of the $V_{zz}$ (shown in Figure~\ref{fig:3-results-VzzOrientation}), defined by the tilt angle $(\theta)$ from the $c$-axis [001] and the azimuthal angle $(\phi)$ from the $a$-axis [100], can help understand the charge distribution further. For the P4/nmm phase, both $\theta$ and $\phi$ remain at $0^{\circ}$ across all cation sites. This was expected from the $\eta$ values discussed previously, it confirms the axial orientation of the electronic gradient along the $c$-axis, possibly linked to the vertical stretching of the lattice and the Ti-ion displacement towards the apical oxygen.

In the lower symmetry phases, the Na-site $V_{zz}$ remains appoximately colinear with [001], with $\theta < 5^{\circ}$. In contrast, the Ti and R sites show significant angular reorientations that correlate precisely with peaks in the $\eta (r_{i})$ trends seen in Figure~\ref{fig:3-Results-DFT-EFGs/VZZ_ETA_RADIUS}. At the Ti site the Pbcm symmetry shows an abrupt shift, where $\theta$ drops from $45^\circ$ to $10^\circ$ between Sm and Pm, with a similar behavior seen for the azimuthal angle in P$\bar{4}$2$_{1}$m close to Tb, where it rises close to $70^\circ$. Conversely, at the R-site the $V_{zz}$ vector in Pbcm flips from a equatorial orientation ($\theta = 90^\circ$) at small ions to a more vertical one ($\theta = 20^\circ$) at the Gd-Eu boundary, coupled with an azimuthal shift towards $45^\circ$, with a similar behaviour seen between Er-Ho for P$\bar{4}$2$_{1}$m where a decrease in tilt and an increase in azimuthal angle is also observed, with the increase in ionic radius, matching the peaks seen in $\eta(r_{i})$. In general, the decrease in tilt of the principal component with the increase in ionic radii corroborates the previous conclusion that the increase in ionic radius favors the stretching of the octahedra rather than their tilting.

As for the Yttrium outlier, its orientation deviates slightly in tilt from the Ln trend at the Ti-site, while maintaining its azimuthal angle, however, at the R-site, Y aligns with the rest of the series, both for $\theta$ and $\phi$. This consistency contrasts with the previously discussed values of $V_{zz}$ where Yttrium exhibits values significantly lower than its neighbors. Given Equation \ref{etadefinition}, the reduced denominator would typically lead to a higher $\eta$, assuming that the numerator behaves similarly across the series. While this is observed for P4/nmm ($\eta = 0$) and Pbcm, the P$\bar{4}$2$_{1}$m is distinct in the fact that $\eta$ for Yttrium is smaller than its peers. This suggests that even though $V_{zz}$ is much smaller than the other Lanthanides with similar ionic radius, the $V_{xx}-V_{yy}$ term, which stands for the planar anisotropy, is sufficiently small in NaYTiO$_{4}$ that is overcompensates the reduced principal component. Consequently, this might be happening to a degree in the other symmetries, though not enough for $\eta$ to be smaller than the neighboring rare-earths. Even so, this leads us to the conclusion that while the orientation of the principal component at the Y site follows the lanthanide trend, the charge distribution in the $a$ and $b$ directions seems much more isotropic ($V_{xx}\approx V_{yy}$) than the other rare-earth sites.

\subsubsection{Electronic Structure and Bandgap evolution}
\label{sec:BandgapPDOSElf}

\begin{figure}[b]
	\centering
	\includegraphics[width=1\columnwidth]{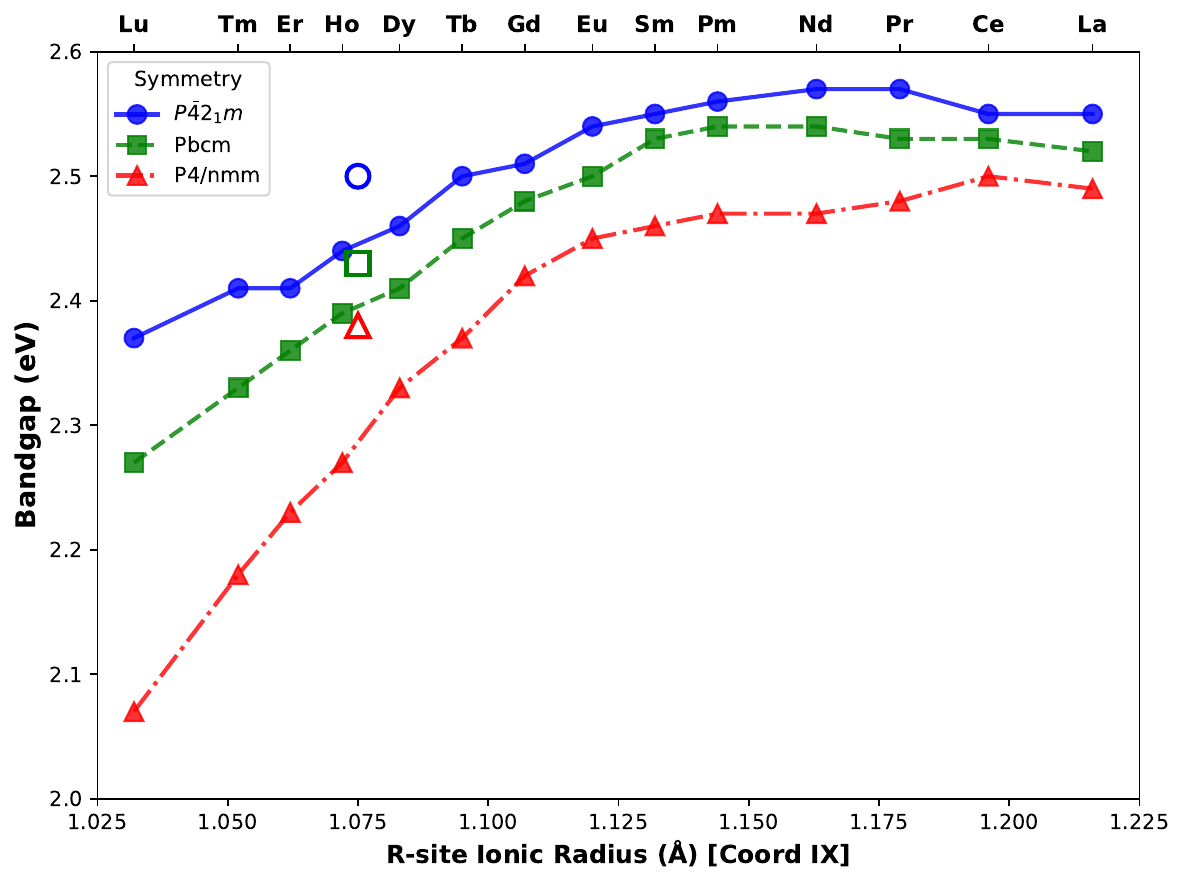}
	\caption{%
		DFT-calculated values for the bandgap of each NaRTiO$_{4}$ compound for all three symmetries in study. All three symmetries are highlighted in distinct colors: P4/nmm (red); P$\bar{4}$2$_{1}$m (blue); Pbcm (green). Yttrium data points are represented by hollow markers.
        }
	\label{fig:3-results/Bandgap}
\end{figure}

The systematic evolution of the electronic bandgap provides a macroscopic signature of the structural coordination and behaviours discussed previously, as the energy gap between valence and conduction bands is directly governed by the O $2p$ and Ti $3d$ orbitals, respectively \cite{mukherjee2025first}. As shown in Figure~\ref{fig:3-results/Bandgap}, the bandgap generally increases with the rare-earth ionic radius across all three symmetries. In the small ion limit, the phases show clearly distinct values, with P4/nmm exhibiting the smallest bandgap, followed by the orthorhombic Pbcm and P$\bar{4}$2$_{1}$m, with the rate of increase across the series being seemingly phase dependent, with P4/nmm showing a steepest slope at lower radii when compared with the remaining 2 symmetries. However, consistent with what was previously observed both for structural and EFG results, the bandgap values for all three symmetries plateau and approach similar values for larger cations, with differences between the phases narrowing down to approximately $0.1$ eV. 

Furthermore, consistent with its behaviour in previous discussions, Yttrium remains a notable outlier, showing a bandgap significantly larger than its direct ionic radius neighbors and more comparable to the values found for Gd or Tb. This deviation is likely a consequence of the higher internal octahedral distortion ($\sigma$) and larger octahedral height previously identified for NaYTiO$_{4}$, which results in a shorter Ti-O apical bond, possibly altering the $p$ - $d$ hybridization in such a way that the conduction band states are moved higher in energy, thus widening the gap.

It should also be noted that the current calculations utilize the PBE functional, which is well known to severely underestimate electronic bandgaps compared to hybrid functionals like HSE06\cite{krukau2006influence}. However, as demonstrated by Mukherjee \textit{et al.}\cite{mukherjee2025first}, the primary difference between PBE and HSE seen for NaYTiO$_{4}$ under the Pbcm phase was the rigid shift of the conduction band towards higher energies, leading to an increase is band gap close to 2 eV, without any significant alterations to the dispersion of the energy states. Thus, though the magnitudes of the bandgaps reported here are to be taken as underestimations, the qualitative trends and relative behavior between phases are expected to withstand the change to a hybrid functional.

Finally, in order to investigate the origin of Yttrium's persistent outlier behavior, both projected density of states (PDOS) and electron localization function (ELF) calculations were performed for all structures across the series (see Supplementary Information). We previously hypothesized that Yttrium's 4d 5s valence configuration, distinct from the 5d 6s present in the Lanthanide ions, might be the trigger to its outlier behaviour. However, the PDOS results show that its orbital occupations and hybridizations remain nearly identical to those of its lanthanide neighbors, with the only notable differences being the non-comparable shifts in Fermi energy and previously discussed value of its bandgap.

Similarly, the ELF topology reveals no significant differences in its spatial distribution. These results are particularly striking when contrasted with the EFG data, which suggested a more isotropic planar environment ($V_{xx} \approx V_{yy}$) at the Y site. The absence of a clear electronic signature in either the DOS or ELF indicates that NaYTiO$_{4}$ structural deviations are not accompanied by a fundamental change in the bonding nature. Consequently, the electronic origin of its outlier status remains poorly understood, suggesting that these specific electronic descriptors cannon capture the subtle mechanism behind it.

%\subsection{Analysis on the influence of the Hubbard \textit{U} Parameter}
%\label{sec:resultsU}
%\input{sections/3-resultsAbInitio_U}

%\subsection{PAC Results and Super-cell Calculations}
%\label{sec:resultsPAC_SC}
%\input{sections/3-results-PAC_SuperCell}

\section{Conclusions}
\label{sec:5-conclusions}
In this work, we provided a comprehensive \textit{ab initio} characterization of the structural and electronic behaviour of the NaRTiO$_{4}$ (R = rare-earth) series across the two previously reported ground states, Pbcm and P$\bar{4}$2$_{1}$m, as well as the aristotype, P4/nmm. Our results show that as the rare-earth ionic radius increases the system favors an axial stretching of the TiO$_{6}$ octahedra, reducing the tilt and increasing the distortion of these structures. This transition is then marked by a convergence in both total energy and macroscopic lattice parameters, where the high-symmetry phase P4/nmm eventually emerges as a more favourable symmetry with the increase in ionic radius. We conclude that this axial elongation, represented by the Ti ion displacement towards the apical oxygen, acts as the primary driver for structural stabilization in the large ion limit.

As for the EFG, their analysis showed that the peaks identified in the $\eta(r_{i})$ trends for the Ti and R sites correspond to critical points where the principal component $V_{zz}$ reorients itself, possibly signaling a crossover point from a tilting dominated regime to a distortion dominated one at larger radii. The strictly vertical orientation of the $V_{zz}$, aligned with the $c$-axis, for P4/nmm coupled with the fact that the EFG parameters for all symmetries approach each other in the large ion limit (Ce; La), suggests that the local electronic environment becomes approachingly uniform regardless of symmetry at this stage. This in turn favors the hypothesis that the specific rotational degrees of freedom seen in the ground states at smaller ionic radii are suppressed, or become less advantageous, in favor of a long-axis deformation, allowing for the alignment of $V_{zz}$ with the $c$-axis. Crucially, the distinct evolution of $V_{zz}$ and $\eta$ at the R and Ti sites provide a clear roadmap for experimental distinction of the two candidate groundstates: Pbcm and P$\bar{4}$2$_{1}$m. By mapping these phase specific signatures, this work provides the necessary benchmark through which experimental measurements via techniques such as NMR or PAC can now be performed, with probes such as $^{23}$Na, $^{44}$Ti or $^{172}$Lu, in order to resolve the ground state phase of the NaRTiO$_{4}$ (R = rare-earth) series.   

Finally, while the bandgap evolution reinforces the trend of structural convergence towards favoring octahedral distortion, the specific outlier behavior of NaYTiO$_{4}$ remains an interesting anomaly. Despite exhibiting larger bandgap and a more pronounced distortion factor than its lanthanide counterparts of similar ionic radii, the PDOS and ELF analysis revealed no fundamental difference between Y and the Lanthanide ions that could further help elucidate the origin of this discrepancy. While the EFG analysis hinted at a more isotropic planar environment at the Y site, the lack of any corresponding piece of evidence in the PDOS and ELF data suggests that Yttrium's deviation might be governed by subtle couplings not fully captured by these descriptors and which remain absent in the Lanthanide series. Nevertheless, this work establishes a framework for the future investigation and eventual resolution of this outlier behaviour.

\section*{Acknowledgments}

We acknowledge the support from FCT through the projects, 2024.00223.CERN (funded by the PRR, RE-C06-i06.m02, through EMRP and FCT, \url{https://doi.org/10.54499/2024.00223.CERN}). Additionally, FCT provided funding via the projects UIDB/04968/2025 and UIDP/04968/2025, and LA/P/0095/2020  (\url{https://doi.org/10.54499/la/p/0095/2020}). We also acknowledge financial support from individual grants: PRT/BD/154996/2023 (\url{https://doi.org/10.54499/PRT/BD/154996/2023}),  2022.04845.CEECIND/CP1719/CT0008 (\url{https://doi.org/10.54499/2022.04845.CEECIND/CP1719/CT0008}) 2023.07340.CEECIND/CP2833/CT0006 (\url{https://doi.org/10.54499/2023.07340.CEECIND/CP2833/CT0006}), 2023.01884.BD (\url{https://doi.org/10.54499/2023.01884.BD}).
We acknowledge CNPq and the support from FAPESP (Project No. 2022/10095-8). We also acknowledge Minho Advanced Computing Center through Project No. 2024.09374.CPCA.A2 and 2025.095592.CPCA.A2

% ---------------- Bibliography ----------------
% Using apsrev style (typical for APS). Ensure ./my-refs-PS.bib is up-to-date.
% If you prefer biblatex/biber, adapt this block accordingly.
\bibliographystyle{apsrev}
\bibliography{./my-refs-PS.bib}

% ---------------- Appendix ----------------
% Use LaTeX \appendix to switch to appendix mode and A,B,... numbering.
%\appendix
% We add a label for the appendix block and adjust float numbering to include
% the appendix letter as prefix for figures/tables.
%\appendix\label{sec:appendix}
%\renewcommand\thefigure{\thesection\arabic{figure}}
%\renewcommand\thetable{\thesection\arabic{table}}
%\renewcommand\thesection{\Alph{section}}
%\renewcommand\thesubsection{\thesection.\arabic{subsection}}
%\input{sections/6-appendix}

%% -------------------- SUPPORTING INFORMATION (SI) --------------------
% -------------------- SUPPORTING INFORMATION (SI) --------------------
% The SI is placed after the main document.
% We switch to single-column and reset counters to use "S" numbering (S1, S2,...).
\clearpage
\onecolumngrid
\pagestyle{plain}
\markboth{}{}

% Reset counters for SI and switch numbering styles for sections, figures, etc.
\setcounter{section}{0}
\setcounter{subsection}{0}
\setcounter{figure}{0}
\setcounter{table}{0}
\setcounter{equation}{0}

% Section labels: S1, S2, ...
\renewcommand{\thesection}{S\arabic{section}}
\renewcommand{\thesubsection}{\thesection.\arabic{subsection}}

% Per-section numbering for floats and equations: Figure S<sec>.<fig>
\renewcommand{\thefigure}{\thesection.\arabic{figure}}
\renewcommand{\thetable}{\thesection.\arabic{table}}
\renewcommand{\theequation}{\thesection.\arabic{equation}}

% Reset figure/table counters at each new SI section (so numbering is Section-local)
\makeatletter
\@addtoreset{figure}{section}
\@addtoreset{table}{section}
\@addtoreset{equation}{section}
\makeatother

% SI title block (you can edit or expand as needed)
\begin{center}
  {\LARGE\bfseries Supporting Information}\\[1ex]
\end{center}
\vspace{1em}

\section{Octahedral Tilt as a Function of Ionic Radii and Distortion Factor}
\label{sec:TiltDist}

\vspace*{\fill}

\begin{figure}[H] % [H] forces it exactly here
    \centering
    \includegraphics[width=\textwidth]{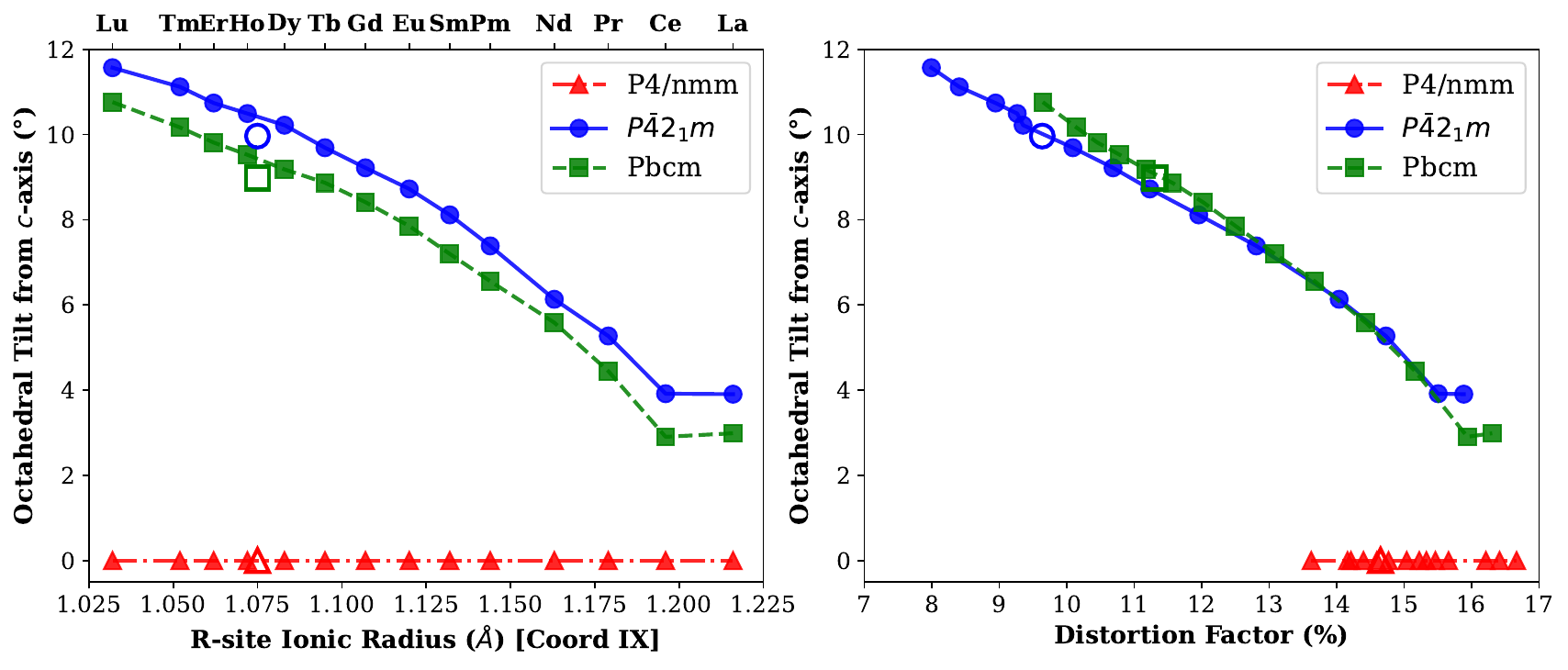}
    \caption{Octahedral Tilt in relation to the $c$-axis in NaRTiO$_{4}$ (R = rare-earth), as functions of R-site ionic radius (left) and distortion factor (right). All three symmetries are plotted: P$\bar{4}$2$_{1}$m (blue), Pbcm (green), P4/nmm (red). Yttrium data points are represented by hollow markers.}
    \label{fig:3-Results/TiltDistortion}
\end{figure}

\vspace*{\fill}

\newpage 
% --- SECTION 1: EFG ---
\section{Electric Field Gradient parameters as functions of octahedral distortion}
\label{sec:EFG_DIST}

\vspace*{\fill}

\begin{figure}[H] % [H] forces it exactly here
    \centering
    \includegraphics[width=\textwidth]{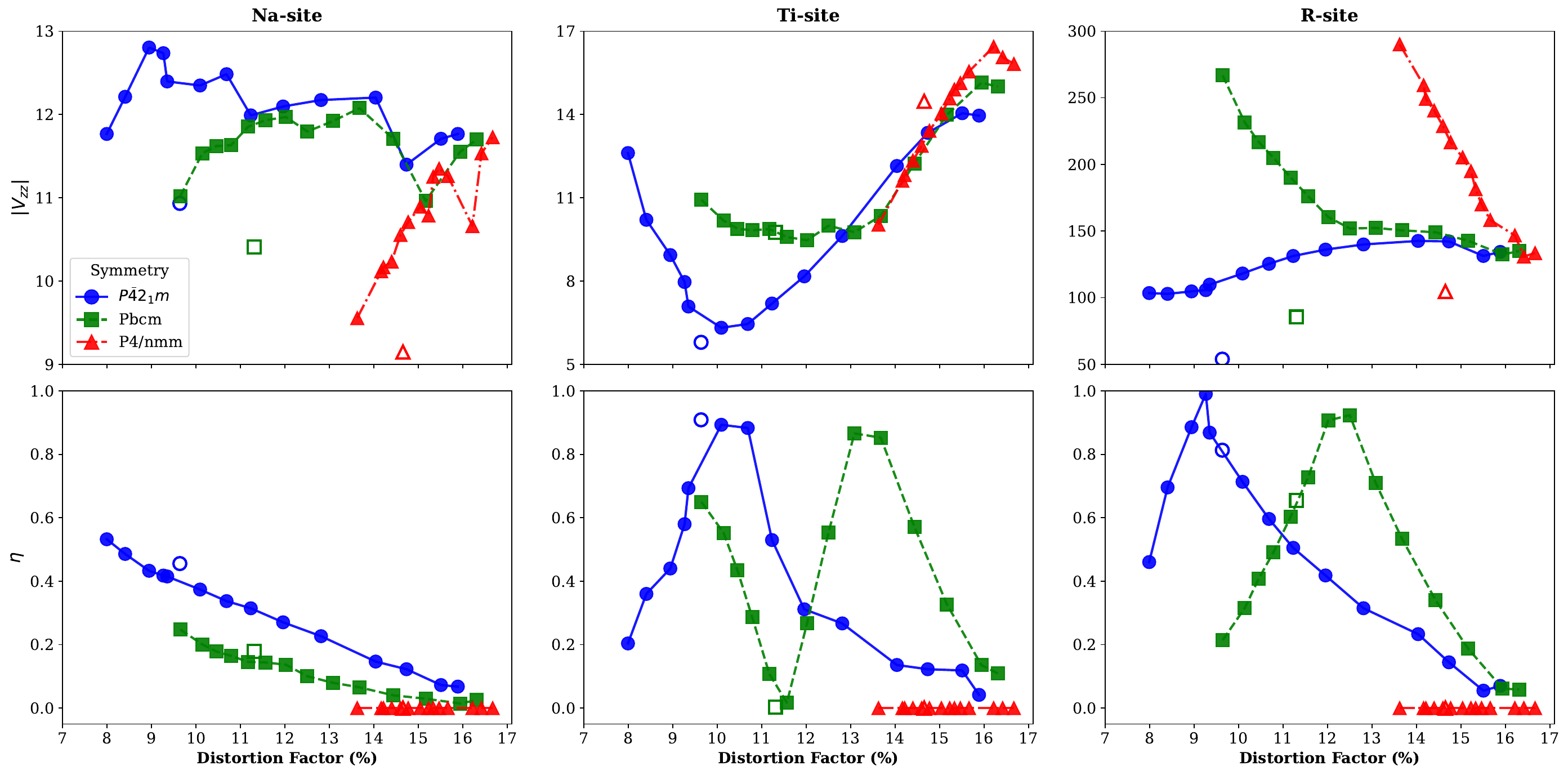}
    \caption{Theoretical electric field gradient (EFG) parameters in NaRTiO$_{4}$ (R = rare-earth), as functions of octahedral distortion, $\sigma$, for different atomic sites: (Left) Na, (Middle) Ti, and (Right) R. The upper panel shows the absolute principal component $|V_{zz}|$ (in $V  \AA^{-2}$), while the lower panel displays the asymmetry parameter $\eta$, both calculated for the three symmetries in study: P$\bar{4}$2$_{1}$m (blue), Pbcm (green), P4/nmm (red). Yttrium data points are represented by hollow markers.}
    \label{fig:3-Results/VzzEtaDistortion}
\end{figure}

\vspace*{\fill}

\newpage

% --- SECTION 2: PDOS ---
\section{Projected Density of States results}
\label{sec:PDOS}

\begin{figure}[H]
    \centering
    \includegraphics[width=0.92\textwidth]{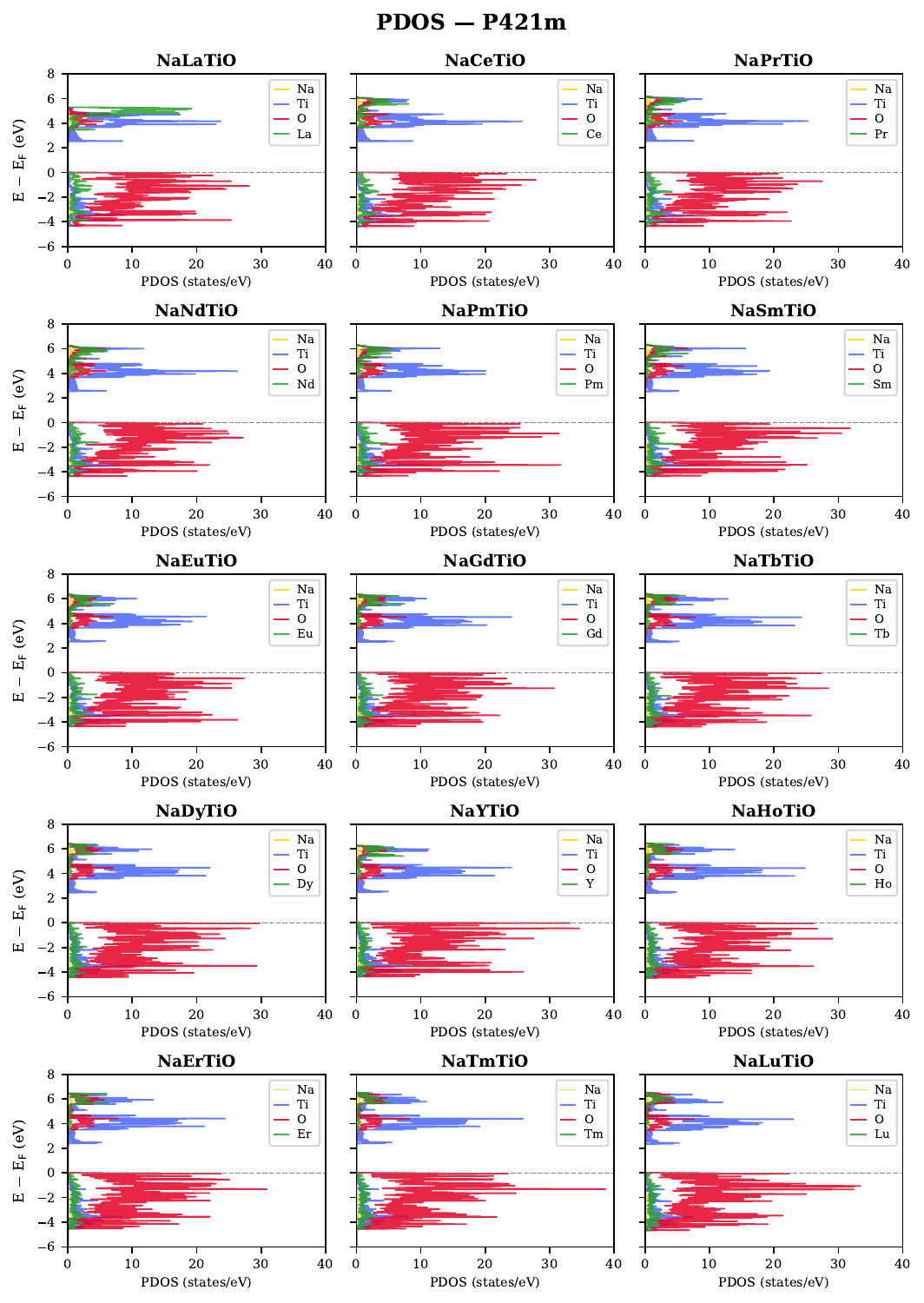}
    \caption{Projected Density of States (PDOS) plots for the P$\bar{4}$2$_{1}$m symmetry across the NaRTiO$_{4}$ symmetry. The orbitals are mapped with the following colors: Na (yellow), Ti (blue), R-site (green); O (red).}
    \label{fig:3-Results/PDOS_P421M}
\end{figure}

\newpage % Move Pbcm to its own page

\begin{figure}[H]
    \centering
    \includegraphics[width=0.92\textwidth]{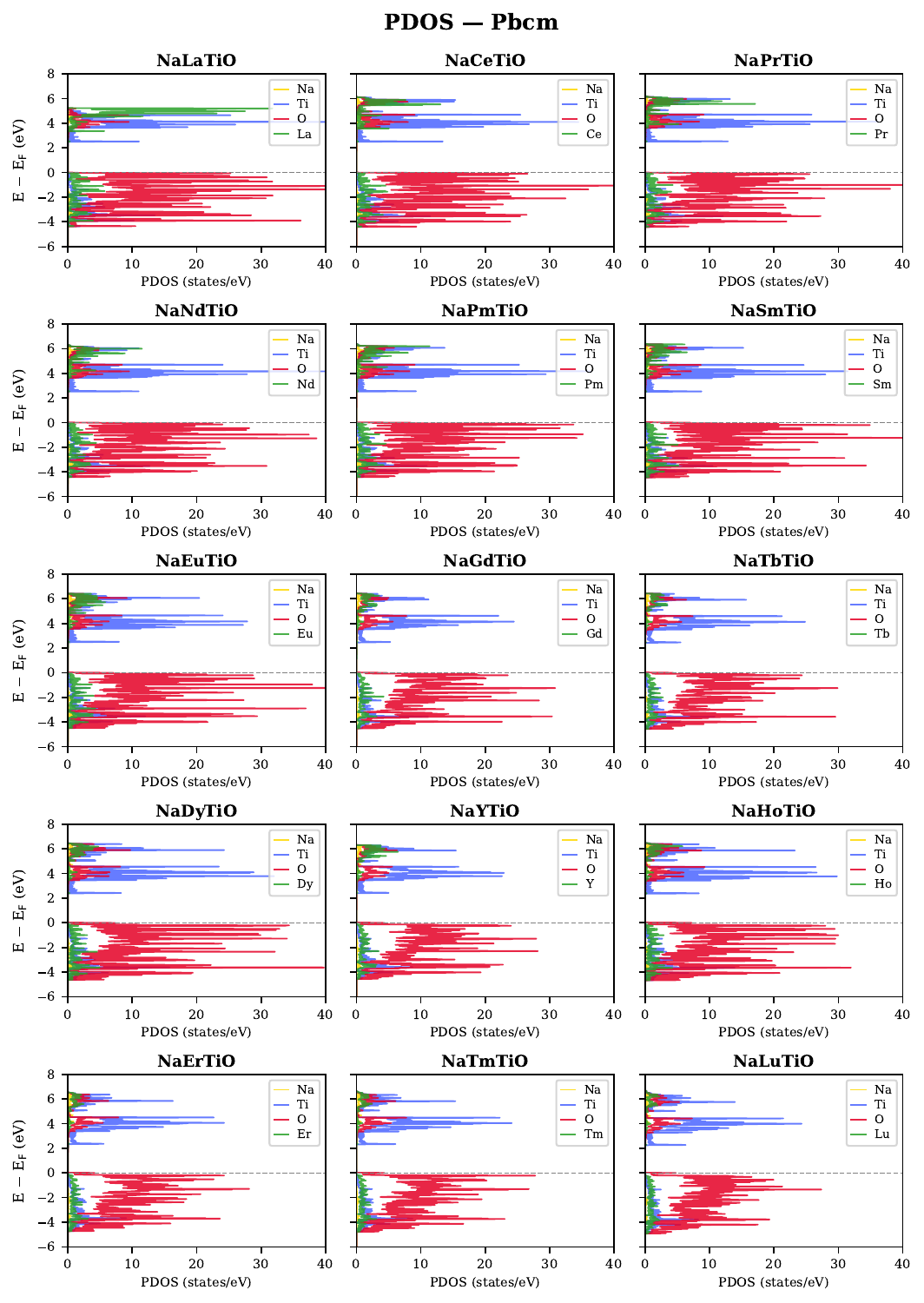}
    \caption{Projected Density of States (PDOS) plots for the Pbcm symmetry across the NaRTiO$_{4}$ symmetry. The orbitals are mapped with the following colors: Na (yellow), Ti (blue), R-site (green); O (red).}
    \label{fig:3-Results/PDOS_PBCM}
\end{figure}

\newpage % Move P4/nmm to its own page

\begin{figure}[H]
    \centering
    \includegraphics[width=0.92\textwidth]{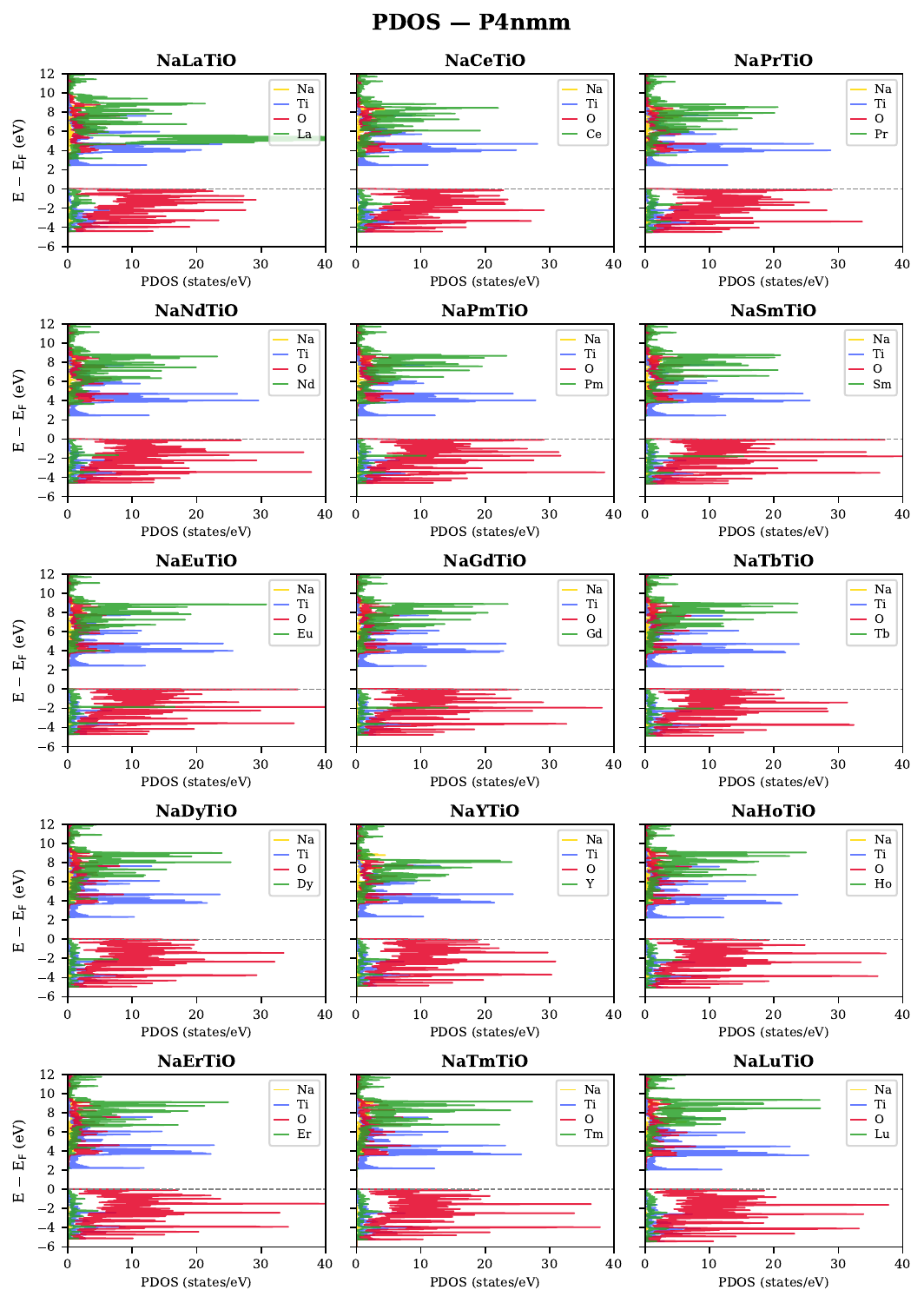}
    \caption{Projected Density of States (PDOS) plots for the P4/nmm symmetry across the NaRTiO$_{4}$ symmetry. The orbitals are mapped with the following colors: Na (yellow), Ti (blue), R-site (green); O (red).}
    \label{fig:3-Results/PDOS_P4NMM}
\end{figure}

\newpage

% --- SECTION 3: ELF ---
\section{Electron Localization Function results}
\label{sec:ELF}

\begin{figure}[H]
    \centering
    \includegraphics[width=\textwidth]{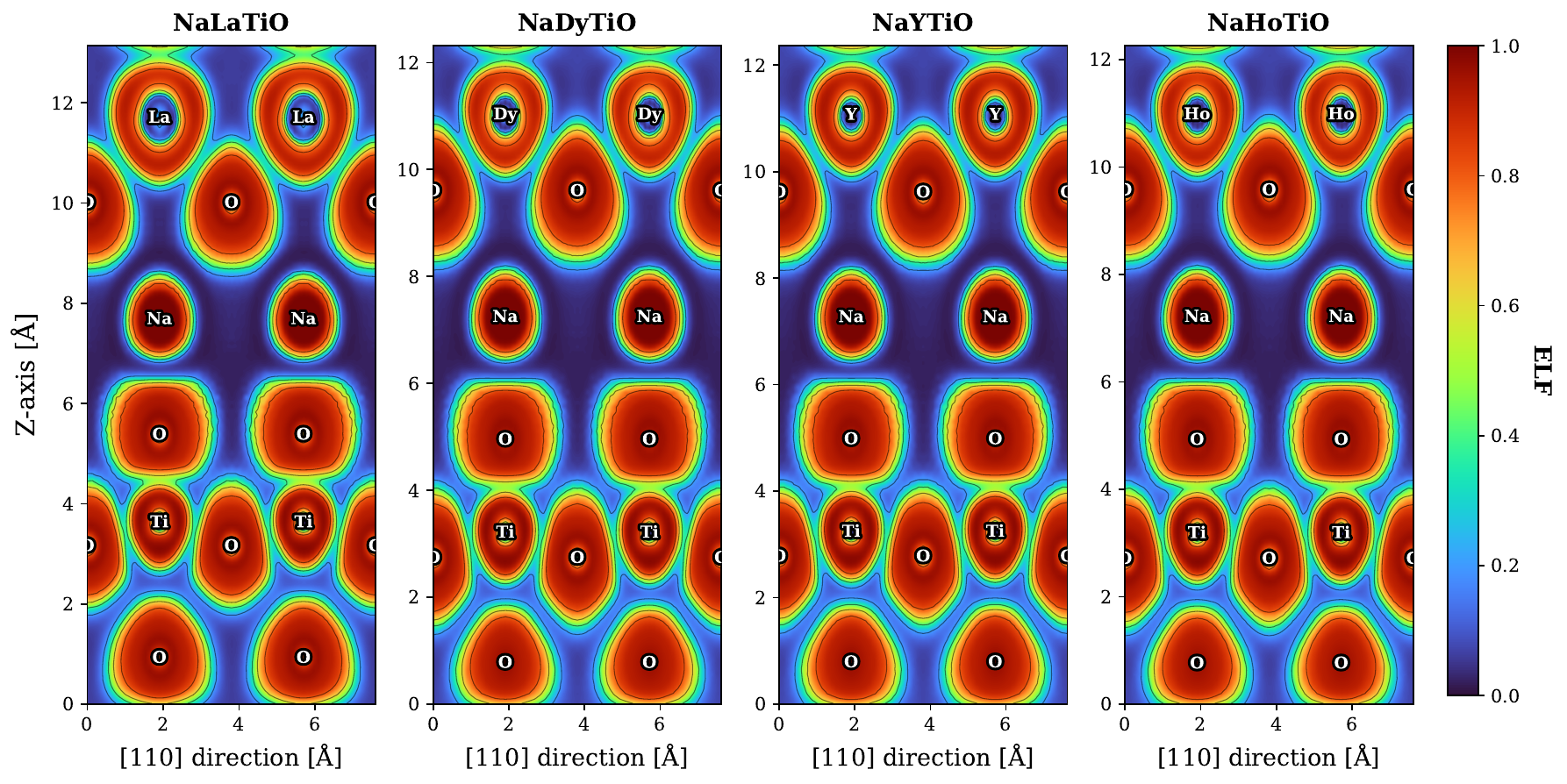}
    \caption{Electron Localization Function (ELF) results of a cut alongside the [-110] direction in the P$\bar{4}$2$_{1}$m phase.}
\end{figure}

\begin{figure}[H]
    \centering
    \includegraphics[width=\textwidth]{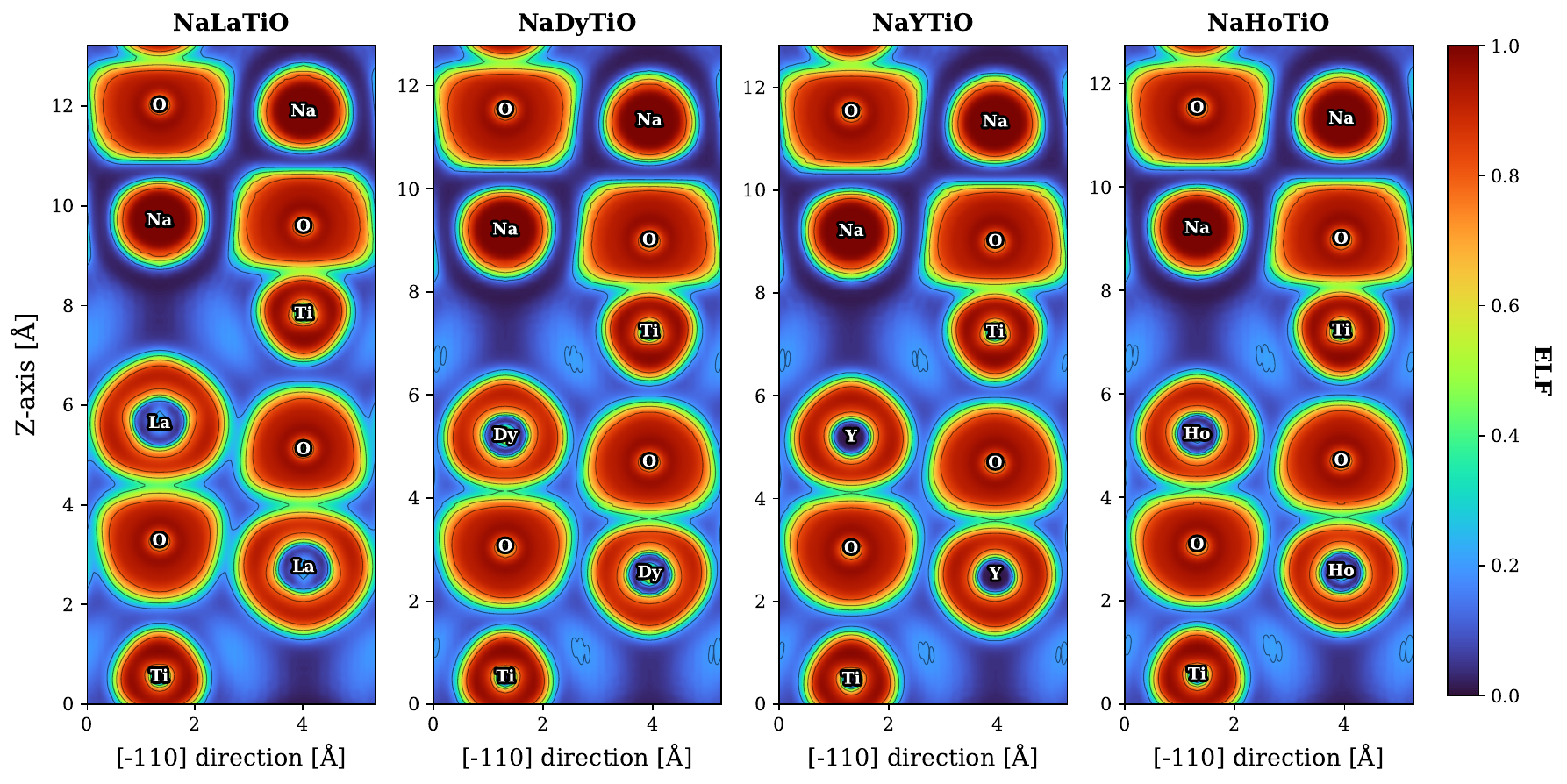}
    \caption{Electron Localization Function (ELF) results of a cut alongside the [-110] direction in the P4/nmm phase.}
\end{figure}
%% ---------------- End of document ----------------
\end{document}